\documentclass[useAMS,usenatbib]{mn2e}
\usepackage{journals}
\usepackage{graphicx}
\usepackage{color}
\usepackage{times}
\usepackage{hyperref}
\title[Supernovae and their host galaxies -- II]{Supernovae and their host galaxies -- II.
The relative frequencies of supernovae types in spirals}
\author[A.~A.~Hakobyan~et~al.]{A.~A.~Hakobyan,$^{1}$\thanks{E-mail:
hakobyan@bao.sci.am}
T.~A.~Nazaryan,$^{1}$
V.~Zh.~Adibekyan,$^{2}$
A.~R.~Petrosian,$^{1}$
\newauthor
L.~S.~Aramyan,$^{1}$
D.~Kunth,$^{3}$
G.~A.~Mamon,$^{3}$
V.~de~Lapparent,$^{3}$
E.~Bertin,$^{3}$
\newauthor
J.~M.~Gomes$^{2}$
and M.~Turatto$^{4}$
\\
$^{1}$Byurakan Astrophysical Observatory, 0213 Byurakan, Aragatsotn province, Armenia\\
$^{2}$Centro de Astrof\'{i}sica da Universidade do Porto, Rua das Estrelas, 4150-762 Porto, Portugal\\
$^{3}$Institut d'Astrophysique de Paris (UMR 7095: CNRS \& UPMC), 98bis Bd Arago, 75014 Paris, France\\
$^{4}$INAF -- Osservatorio Astronomico di Padova, Vicolo dell'Osservatorio 5, 35122 Padova, Italy}
\begin{document}

\date{Accepted 2014 August 5. Received 2014 August 2; in original form 2014 May 27}

\pagerange{\pageref{firstpage}--\pageref{lastpage}} \pubyear{2014}

\maketitle

\label{firstpage}

\begin{abstract}
  We present an analysis of the relative frequencies of different supernova (SN) types in
  spirals with various morphologies and in barred or unbarred galaxies.
  We use a well-defined and homogeneous sample of spiral host galaxies of 692 SNe
  from the Sloan Digital Sky Survey in different stages of galaxy--galaxy interaction and activity
  classes of nucleus. We propose that the underlying mechanisms shaping
  the number ratios of SNe types can be interpreted within the framework of
  interaction-induced star formation, in addition to the known relations between
  morphologies and stellar populations.
  We find a strong trend in behaviour of the $N_{\rm Ia}/N_{\rm CC}$ ratio depending on
  host morphology, such that early spirals include more Type Ia SNe.
  The $N_{\rm Ibc}/N_{\rm II}$ ratio is higher in a broad bin of early-type hosts.
  The $N_{\rm Ia}/N_{\rm CC}$ ratio is nearly constant when changing from normal,
  perturbed to interacting galaxies, then declines in merging galaxies,
  whereas it jumps to the highest value in post-merging/remnant galaxies.
  In contrast, the $N_{\rm Ibc}/N_{\rm II}$ ratio jumps to the highest value in merging
  galaxies and slightly declines in post-merging/remnant subsample.
  The interpretation is that the star formation rates and morphologies of galaxies, which are strongly affected
  in the final stages of interaction, have an impact on the number ratios of SNe types.
  The $N_{\rm Ia}/N_{\rm CC}$ ($N_{\rm Ibc}/N_{\rm II}$) ratio increases (decreases) from star-forming
  to active galactic nuclei (AGN) classes of galaxies. These variations are consistent with the scenario of
  an interaction-triggered starburst evolving into AGN during the later stages of interaction,
  accompanied with the change of star formation and transformation of
  the galaxy morphology into an earlier type.
\end{abstract}


\section{Introduction}

There is a widespread consensus that different kinds of observables of galaxies are
related directly to the different stellar populations and to star formation.
These observables range from the overall morphology, disturbance due to interaction, presence of bar,
activity classes of nucleus and occurrence of supernovae (SNe) of various types.

Different levels of activity
in the central region of galaxies can be powered by either
accretion on to a supermassive black hole
\citep[SMBH; e.g.][]{2010MNRAS.407.1529H},
or combination of SMBH induced activity with a burst of
circumnuclear massive star formation
\citep[e.g.][]{2007AJ....134..648M},
or mostly massive star formation alone
\citep[e.g.][]{2012MNRAS.423.3274D}.
Association of active galactic nuclei (AGN) with circumnuclear
star formation is common in many bright Seyfert (Sy) galaxies
\citep[e.g.][]{2005A&A...429..141K,2009AJ....137.3548P}
and is supported by modelling
of AGN-triggered star formation
\citep[e.g.][]{2009ApJ...700..262S},
and/or the onset of star formation with subsequently fuelled AGN
\citep[e.g.][]{2007ApJ...671.1388D,2012MNRAS.420L...8H}.
The star formation in AGN host galaxies is not concentrated
primarily in the nuclear regions, but is distributed over scales
of at least several kiloparsecs
\citep[e.g.][]{2003MNRAS.346.1055K,2007ApJS..173..357K}.

A possible mechanism to explain the co-evolution of an AGN and its
host galaxy is that both nuclear and extended activity are
triggered by interacting/merging of galaxies
(e.g. \citealt{2001ApJ...559..147S,2008ApJ...679.1047K};
\citealt*{2013MNRAS.430..638S}).
The produced gas inflow not only forms massive stars in the central region,
but also fuels the SMBH in the centre of galaxies
(e.g. \citealt*{2005Natur.433..604D}).
There also exists the possibility, depending on accretion rate,
that enhanced star formation in discs does not correlate with the nuclear
activity but with interaction
\citep[e.g.][]{2006AJ....131..261J,2012MNRAS.423.3274D}.
An additional possibility is that gravitational instabilities in the disc
of barred galaxies can cause gas transfer to the central regions of galaxies
\citep[e.g.][]{2004ApJ...612L..17W}.
In this respect, the presence of bars can play
an important role in the star formation of galaxies
\citep[e.g.][]{2011MNRAS.416.2182E}.

The triggered bursts of star formation strongly affect the observed numbers of SNe
(e.g. \citealt*{2002MNRAS.331L..25B}; \citealt{2005AJ....129.1369P};
\citealt*{2012MNRAS.424.2841H}; \citealt{2013MNRAS.436.3464K})
in addition to the known relations
between the rate of various SN types and
the stellar content of galaxies with different morphologies
(e.g. \citealt*{1999A&A...351..459C};
\citealt{2005A&A...433..807M,2011MNRAS.412.1473L}).
Core-Collapse (CC) SNe, whose progenitors are thought to be young massive stars
\citep[e.g.][]{2003LNP...598...21T,
2009ARA&A..47...63S,2012MNRAS.424.1372A},
are observationally classified in three major classes, according to
the strength of lines in optical spectra \citep[e.g.][]{1997ARA&A..35..309F}:
Type II SNe show hydrogen lines in
their spectra, while Types Ib and Ic do not, with Type Ib SNe showing helium
and Type Ic SNe showing neither hydrogen nor helium.
There are two proposed channels for stripping the hydrogen and helium envelopes:
(1) massive Wolf--Rayet stars with large ($M \geq 30~M_{\odot}$) main-sequence masses that
experience strong mass-loss \citep[e.g.][]{2003ApJ...591..288H};
(2) lower mass binaries stripped through interaction by a close companion
(e.g. \citealt*{1992ApJ...391..246P}).
Therefore, the progenitors of SN~Ibc\footnote{{\footnotesize By SN~Ibc, we
denote stripped-envelope SNe of Type Ib, Ic, and mixed Ib/c
whose specific subclassification is uncertain.}}
may be more massive than those of normal SN~II ($8-16~M_{\odot}$; \citealt{2009ARA&A..47...63S})
or explode in regions of higher metallicity \citep[e.g.][]{2003ApJ...591..288H}.
Due to short-lived massive progenitors
($<0.1$~Gyr; \citealt{1992A&AS...96..269S}),
CC SNe are considered as good tracers of recent star formation in galaxies
(e.g. \citealt{1990A&A...239...63P,1995A&A...297...49P};
\citealt{1999A&A...351..459C};
\citealt{2005AJ....129.1369P,2009MNRAS.399..559A,
2009A&A...508.1259H,2012MNRAS.424.1372A,2013MNRAS.428.1927C,
2013MNRAS.436.3464K,2014ApJ...791...57S}).
Thermonuclear SNe, alias Type Ia SNe, result from stars of different ages
($>0.5$~Gyr), with even the shortest lifetime progenitors
having much longer lifetime than the progenitors of CC SNe \citep[e.g.][]{2012PASA...29..447M}.
Their rates are described as a sum of two terms, one depending on
the current star formation rate (SFR) and another on the total stellar mass
\citep[e.g.][]{2005A&A...433..807M,2005ApJ...629L..85S,
2011Ap.....54..301H,2011MNRAS.412.1473L}.
Type Ia SNe are associated less tightly to star formation in spirals and irregulars
\citep[e.g.][]{2006A&A...453...57J}.

Several authors have studied the number ratios
\citep[e.g.][]{2003A&A...406..259P,2009A&A...503..137B}
and spatial distributions
(e.g. \citealt*{1992A&A...264..428B};
\citealt{1997AJ....113..197V,2008MNRAS.388L..74F,
2009MNRAS.399..559A,2009A&A...508.1259H})
of different SN types in large numbers of galaxies.
None of these studies have attempted to categorize the hosts according
to their activity classes of nucleus and interaction.
Nevertheless, other authors have shown that
the number ratios
\citep[e.g.][]{2002MNRAS.331L..25B,2005AJ....129.1369P,2014ApJ...791...57S}
and distributions of SNe
(e.g. \citealt{1990A&A...239...63P,2005AJ....129.1369P,
2008Ap.....51...69H}; \citealt*{2010MNRAS.405.2529W};
\citealt{2013MNRAS.436.3464K})
might be different in galaxies with varying activity classes.
For example, \cite{2002MNRAS.331L..25B} showed that the ratio
$N_{\rm Ibc}/N_{\rm II}$ between the total numbers of
SNe~Ibc and SNe~II reflects metallicity,
age, fraction of binary systems, and initial mass function (IMF) shape,
which might be quite different in galaxies with
various activity classes of nuclei.
They also found that the ratio $N_{\rm Ibc}/N_{\rm II}$ measured
in Sy galaxies exceeds that in normal host galaxies by a factor of 4.
Studying a sample of CC SNe in galaxies hosting AGN,
\citet{2005AJ....129.1369P} and \citet{2008Ap.....51...69H} found
that the SNe in active/star-forming (SF) galaxies are more centrally
concentrated than those in normal galaxies.
\citet*{2012A&A...540L...5H} modelled the radial distribution of SNe in the nuclear
starbursts of M~82, Arp~220, and Arp~299A galaxies, and
interpreted the results as evidence of galaxy--galaxy
interactions that are expected to trigger massive
star formation down to the central kiloparsec region of galaxies.
In addition, \citet{2012MNRAS.424.2841H}
presented the results of a reanalysis of
\citet*{2010ApJ...717..342H} with increased statistics and
found a remarkable excess of
CC SNe within the central regions of disturbed galaxies,
i.e., in galaxies showing signs of merger-triggered starbursts in the nuclei.

The locations of SNe in multiple systems of galaxies have also been studied
\citep[e.g.][]{1995A&A...297...49P,2001MNRAS.328.1181N,2013Ap&SS.347..365N}.
There is an indication that the SN rate is higher in galaxy pairs compared
with that in groups \citep{2001MNRAS.328.1181N}.
In addition, SNe Ibc are located
in pairs interacting more strongly
than pairs containing SNe Ia and II
\citep{2013Ap&SS.347..365N}.
These results are considered to be related to the higher SFR
in strongly interacting systems.
\defcitealias{2012A&A...544A..81H}{Paper~I}

However, the aforementioned studies suffer from poor statistics,
as well as strong biases in the SNe and their host galaxies samples.
In our first paper of this series
\citep[][hereafter Paper~I]{2012A&A...544A..81H},
we have created a large and
well-defined data base that combines extensive new measurements and
a literature search of 3876 SNe and their 3679 host galaxies
located in the sky area covered by
the Sloan Digital Sky Survey (SDSS) Data Release 8 (DR8).
This data base is much larger than previous ones,
and is expected to provide a homogeneous
set of global parameters of SN hosts, including morphological
classifications and measures of activity classes of nuclei.
In addition, we have analysed and discussed many selection effects and biases,
which usually affect the studies of SNe.
For more details, the reader is referred to \citetalias{2012A&A...544A..81H}.

In this paper, we investigate the correlations between SNe number ratios
($N_{\rm Ia}/N_{\rm CC}$,
$N_{\rm Ibc}/N_{\rm II}$, and
$N_{\rm Ic}/N_{\rm Ib}$)
and other observable parameters of host galaxies.
We present an analysis of the SNe number ratios in spiral galaxies
with different morphologies and with or without bars.
In addition, we use a well-defined
and homogeneous sample of host galaxies with various levels
of interaction to explore the numbers of SNe resulting from star formation
in their host galaxies as a function of morphological disturbances.
Furthermore, we perform a statistical study of SNe discovered in galaxies
with different activity classes of nuclei
[e.g. SF, composite (C),
low-ionization nuclear emission-line region (LINER), or Sy]
using relative frequencies of the various SN types.
We reveal and discuss the underlying mechanisms
shaping the number ratios of SNe types
within the framework of interaction-induced star formation,
in addition to the well-known relations between morphologies and stellar populations.

\begin{table}
  \centering
  \begin{minipage}{66mm}
  \caption{Types of 14 SNe unclassified in \citetalias{2012A&A...544A..81H}.}
    \label{SN-new-types}
    \begin{tabular}{lll}
    \hline
  \multicolumn{1}{c}{SN name}&\multicolumn{1}{c}{Type$^a$}
  &\multicolumn{1}{c}{Ref.}\\
  \hline
    1988E & II pec~: & \citet{1988IAUC.4546....1M} \\
    1988ab & Ia~: & \citet{1998PASP..110..553R} \\
    1988ac & Ib/c~? & \citet{1998PASP..110..553R} \\
    1992bf & Ia pec~: & \citet{2008ApJ...681..482C} \\
    1992bu & Ib/c~? & \citet{2011MNRAS.416..567A} \\
    1993R & Ia pec~: & \citet{1993IAUC.5842....2F} \\
    1999gs & Ia~: & \citet{2013ApJ...778..167F} \\
    2000ft & II & \citet{2006ApJ...638..938A} \\
    2002bp & Ia pec~: & \citet{2012MNRAS.425.1789S} \\
    2004cs & Ia pec~: & \citet{2013ApJ...767...57F} \\
    2005co & Ia & \citet{2005ATel..532....1B} \\
    2008hl & Ib/c~? & \citet{2013ApJ...778..167F} \\
    2009az & IIP & \citet{2013PhDT.........2K} \\
    2010cr & Ia~: & \citet{2013ApJ...779...23M} \\
  \hline \\
  \end{tabular}
  \parbox{\hsize}{
   $^a$SNe~1988ac and 2008hl are classified as CC SNe in
   \cite{1998PASP..110..553R} and \cite{2013ApJ...778..167F}, respectively.
   However, taking into account the local environments
   \citep[see][and references therein]{2013MNRAS.428.1927C}
   of this SNe, we list them as probable Type Ib/c.}
  \end{minipage}
\end{table}

This is the second paper of a series and
the outline is as follows. Section~\ref{sample} introduces
sample selection.
In Section~\ref{Resdiscus}, we give the results and discuss
all the statistical relations.
Our conclusions are summarized in
Section~\ref{concl}.
Throughout this paper, we adopt a cosmological model with
$\Omega_{\rm m}=0.27$, $\Omega_{\rm \Lambda}=0.73$, and a Hubble constant is taken as
$H_0=73 \,\rm km \,s^{-1} \,Mpc^{-1}$ \citep{2007ApJS..170..377S},
both to conform to the values used in our data base.

\section[]{Sample selection}
\label{sample}

\begin{table}
  \centering
  \begin{minipage}{52mm}
  \caption{New morphological types of 20 host galaxies, which
           were not attributed precise Hubble subclasses in
           \citetalias{2012A&A...544A..81H},
           but were simply labelled as `S'.}
    \label{hosts-new-types}
    \begin{tabular}{lll}
    \hline
  \multicolumn{1}{c}{Galaxy$^a$}&\multicolumn{1}{c}{Morph.$^b$}
  &\multicolumn{1}{c}{Bar}\\
  \hline
    J001911.00+150622.7 & Sm~? & B \\
    J005740.42+434732.0 & Sc~: & B \\
    J021030.84+022005.7 & Sb~: &  \\
    J030211.42--010959.3 & Sm~? &  \\
    J030258.52--144904.4 & Sm~? &  \\
    J043400.04--083445.0 & Sbc~: &  \\
    J073656.63+351431.9 & Sc~ &  \\
    J074726.42+265532.5 & Sm~? &  \\
    J083824.00+254516.3 & Sbc~: &  \\
    J094315.30+361707.1 & Sdm~: &  \\
    J095100.34+200420.4 & Sdm~: &  \\
    J105846.90+592912.5 & Sb~? &  \\
    J112830.77+583342.9 & Sc~? &  \\
    J112833.38+583346.4 & Sc~? &  \\
    J115913.13--013616.0 & Sm~: &  \\
    J145407.71+423253.1 & Sm~: &  \\
    J221930.27+292316.9 & Sc~: &  \\
    J231925.09+055421.6 & Sb~: &  \\
    J235125.02+200641.9 & Sbc~: & B \\
    PGC~71868$^c$ & Sb~: &  \\
  \hline \\
  \end{tabular}
  \parbox{\hsize}{
   $^a$Host galaxy SDSS designation.\\
   $^b$Symbol `:' indicates that the classification is doubtful
   and `?' indicates that the classification is highly uncertain.\\
   $^c$An alternative name is mentioned since there is no photometric object
   in the SDSS at the position of the galaxy.}
  \end{minipage}
\end{table}

The current investigation is based upon the total sample of SNe
and their host galaxies presented in
\citetalias{2012A&A...544A..81H}\footnote{{\footnotesize The parameters
of several SNe and their host galaxies were revised
in \citet{2013Ap.....56..153A}.}}.
This data base contains spectroscopic classes,
accurate coordinates, and offsets from galaxy nucleus
of 3876 SNe (72 SNe~I, 1990 SNe~Ia, 234 SNe~Ibc,
870 SNe~II\footnote{{\footnotesize Note that in SN~II, we include Types IIP, IIL, IIb, and IIn.}},
and 710 unclassified SNe)\footnote{{\footnotesize All the uncertain (`:' or `?') and
peculiar (`pec') classifications are flagged
in table~7 of \citetalias{2012A&A...544A..81H}.
Types I, Ia, and II include also a few SNe classified
from the light curve only.}}
from the coverage of SDSS DR8.
The last SN included in the total sample of \citetalias{2012A&A...544A..81H}
is SN~2011bl \citep{2011CBET.2694....1N}, discovered on 2011 April~5.
Using the SDSS multiband images, photometric, and spectral data,
\citetalias{2012A&A...544A..81H} also provides accurate coordinates,
heliocentric redshifts, morphological types, activity classes of nuclei,
apparent $g$-band magnitudes, major axes $(D_{25})$, axial ratios $(b/a)$,
and position angles of the SNe host galaxies.
In addition, it is noted whether a host has a bar, a disturbed disc,
or is part of an interacting or merging system.

In \citetalias{2012A&A...544A..81H}, we have shown that
the total sample of SNe is largely incomplete beyond ${\rm 100 \pm 3~Mpc}$:
the distributions of SNe of Types~Ibc and II are similar and display
a sharp decline beyond this value
(see fig.~14 in \citetalias{2012A&A...544A..81H}).
Type Ia SNe, because of their comparatively high luminosity and
the presence of dedicated surveys,
are discovered at much greater distances than CC~SNe.
Thus, to avoid biasing the current sample against or in favour of one of
the SN types, we truncate the sample to distances $\leq97~{\rm Mpc}$.

An additional restriction to SNe discoveries is needed,
because Type Ibc SNe were labelled as `I pec' types
during observations before 1986.
Therefore, we also limit the sample to SNe discovered
since 1986. It is important to take into account the fact that
SNe Ia explode in all morphological types of galaxies,
while CC SNe explode mostly in spiral and irregular hosts
(e.g. \citealt*{2005PASP..117..773V};
\citealt{2008A&A...488..523H,2012A&A...544A..81H}).
Therefore, to exclude SNe type biasing due to host-galaxy morphology,
we further restrict the sample to SNe detected in
spiral galaxies (including S0/a types).
Note that irregular hosts are not selected,
because they show neither prominent nuclei nor clear discs,
on which we focus in this article.

After these restrictions, we look through the current literature
for available information on the spectroscopic types of
the 31 unclassified SNe of \citetalias{2012A&A...544A..81H}.
Table~\ref{SN-new-types} presents the collected types for 14 SNe
with the respective references.
In addition, \citet{2011MNRAS.412.1419L}
have reported that there was no SN at the position of
SN~2003dl. Therefore, we remove SN~2003dl from our analysis.
The remaining 16 SNe without spectroscopic classification
(1989ac, 1991bk, 1997bm, 1998cf, 1998cp,
2004ad, 2005cd, 2006A, 2007cd, 2007gq,
2007kc, 2008iv, 2009gt, 2010dh, 2010ha, and 2010lo)
are also removed from the sample.

\begin{figure}
  \begin{center}
  \includegraphics[width=\hsize]{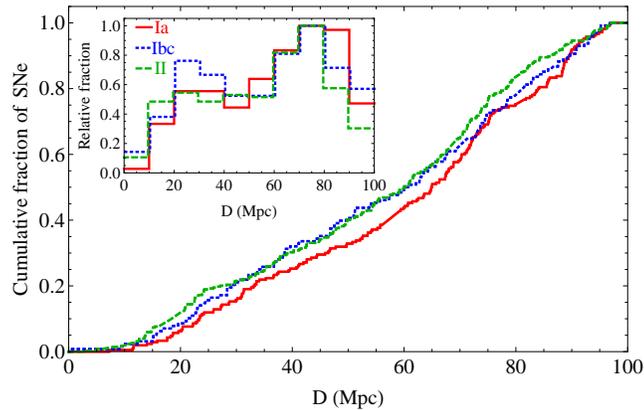}
  \end{center}
  \caption{Cumulative and relative (inset) fractions
  of 692 SNe (Ia -- red solid,
  Ibc -- blue dotted, and II -- green dashed) in S0/a -- Sm galaxies
  as a function of distance, discovered in the time interval between
  1986 January and 2011 April.}
  \label{dSNtype}
\end{figure}

After these operations, we are left with a sample of
692 SNe within 608 host galaxies.
In addition, we morphologically classify
20 host galaxies of 27 SNe, which were without precise Hubble subclasses
(labelled as `S') in \citetalias{2012A&A...544A..81H}.
The results are presented in Table~\ref{hosts-new-types}.
Table~\ref{SN-morph-table} displays the distribution of all SNe types
among the various considered spiral types for their host galaxy.
In Table~\ref{bar-morph}, we also present the distribution of
SNe in barred or unbarred galaxies, for each spiral type.

\begin{table}
  \centering
  \begin{minipage}{78mm}
  \caption{Numbers of SNe discovered since 1986,
  and with a distance $\leq97~{\rm Mpc}$,
  according to morphological type of host galaxies (S0/a -- Sm).
  Among all the SNe types, there are 40 uncertain
  (`:' or `?') and 49 peculiar (`pec') classifications.}
  \tabcolsep 3.5pt
  \label{SN-morph-table}
    \begin{tabular}{lrrrrrrrrrrr}
    \hline
  &\multicolumn{1}{c}{S0/a}&\multicolumn{1}{c}{Sa}
  &\multicolumn{1}{c}{Sab}&\multicolumn{1}{c}{Sb}&\multicolumn{1}{c}{Sbc}&\multicolumn{1}{c}{Sc}&\multicolumn{1}{c}{Scd}
  &\multicolumn{1}{c}{Sd}&\multicolumn{1}{c}{Sdm}&\multicolumn{1}{c}{Sm}&\multicolumn{1}{r}{All}\\
  \hline
    Ia & 27 & 14 & 20 & 33 & 40 & 46 & 13 & 13 & 3 & 1 & 210 \\
    Ib & 0 & 0 & 1 & 5 & 6 & 14 & 4 & 0 & 2 & 0 & 32 \\
    Ib/c & 1 & 3 & 1 & 3 & 6 & 9 & 1 & 2 & 1 & 1 & 28 \\
    Ic & 0 & 1 & 1 & 9 & 22 & 17 & 6 & 8 & 1 & 3 & 68 \\
    II & 2 & 7 & 7 & 41 & 75 & 125 & 35 & 32 & 18 & 12 & 354 \\
  \hline
    All & 30 & 25 & 30 & 91 & 149 & 211 & 59 & 55 & 25 & 17 & 692 \\
  \hline \\
  \end{tabular}
  \end{minipage}
\end{table}

Fig.~\ref{dSNtype} shows the
cumulative and relative fractions
of the 692 SNe with Types Ia, Ibc, and II as a function of distance.
We underline that SNe classified as Ib/c are included in the computation of
$N_{\rm Ibc}$, but are omitted when computing $N_{\rm Ib}$ and $N_{\rm Ic}$;
and SNe classified as IIb and IIn are included in the computation of $N_{\rm II}$.
The Kolmogorov--Smirnov (KS) goodness-of-fit test
shows that the distributions in Fig.~\ref{dSNtype}
are not significantly different from
each other ($P_{\rm KS}^{\rm Ia,Ibc}=0.35$, $P_{\rm KS}^{\rm Ia,II}=0.14$,
$P_{\rm KS}^{\rm Ia,CC}=0.20$, $P_{\rm KS}^{\rm Ibc,II}=0.88$,
and $P_{\rm KS}^{\rm Ib,Ic}=0.63$)
and could thus be drawn from the same parent distribution of distances.
Therefore, our samples of SNe and their host galaxies should not be strongly
affected by any of the redshift-dependent bias discussed in \citetalias{2012A&A...544A..81H}.

\begin{figure*}
\begin{center}$
\begin{array}{@{\hspace{0mm}}c@{\hspace{0mm}}c@{\hspace{0mm}}c@{\hspace{0mm}}c@{\hspace{0mm}}c@{\hspace{0mm}}}
\includegraphics[width=0.2\hsize]{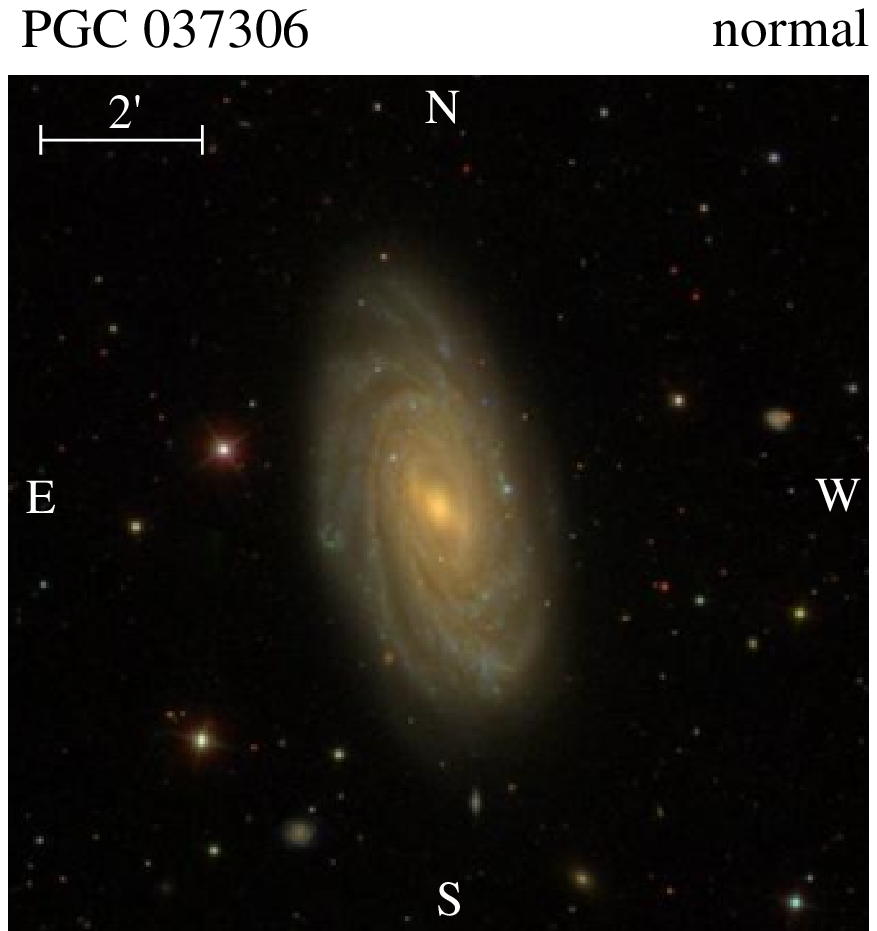} &
\includegraphics[width=0.2\hsize]{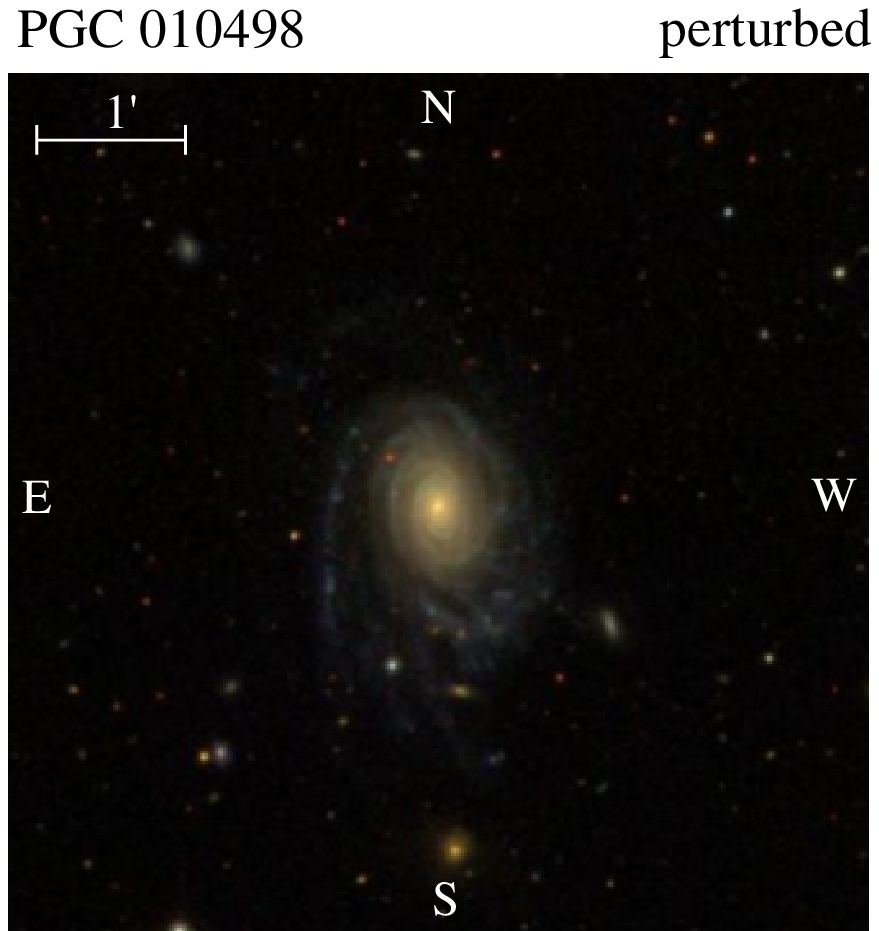} &
\includegraphics[width=0.2\hsize]{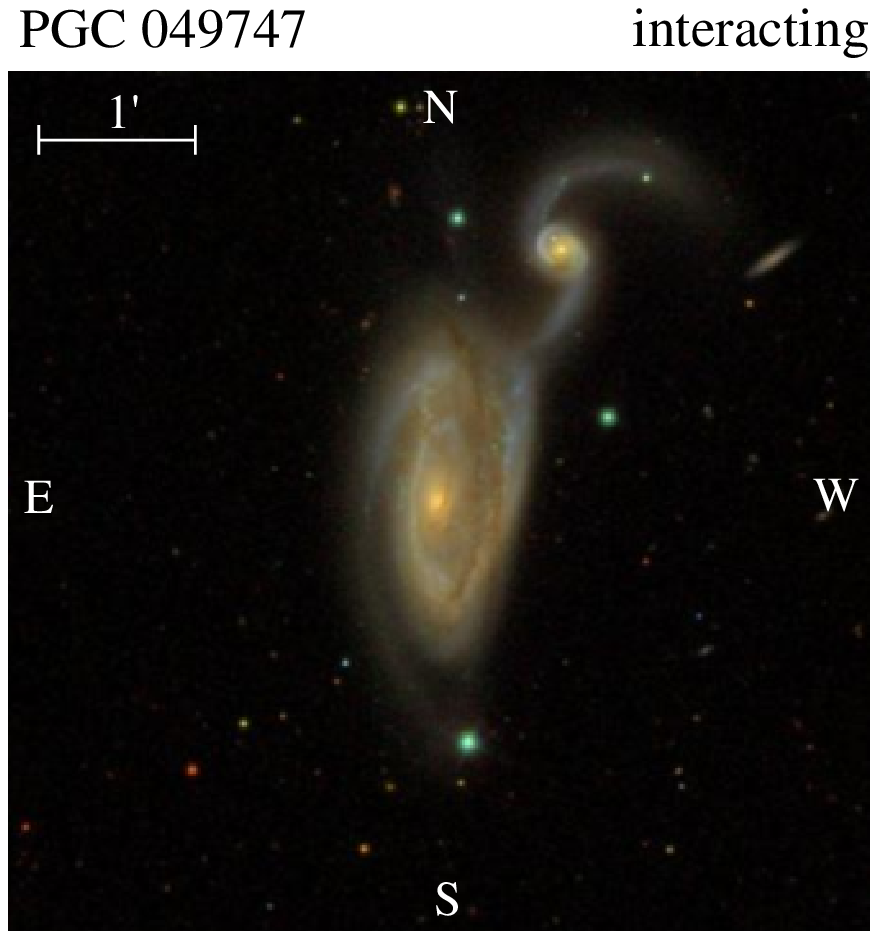} &
\includegraphics[width=0.2\hsize]{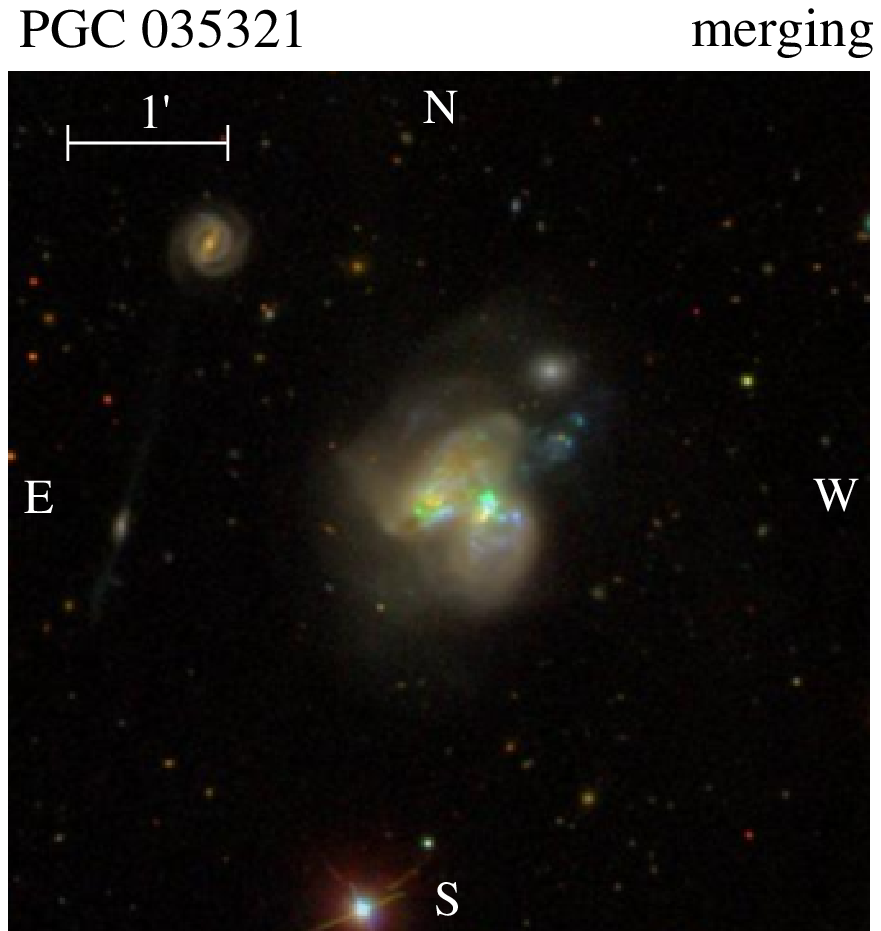} &
\includegraphics[width=0.2\hsize]{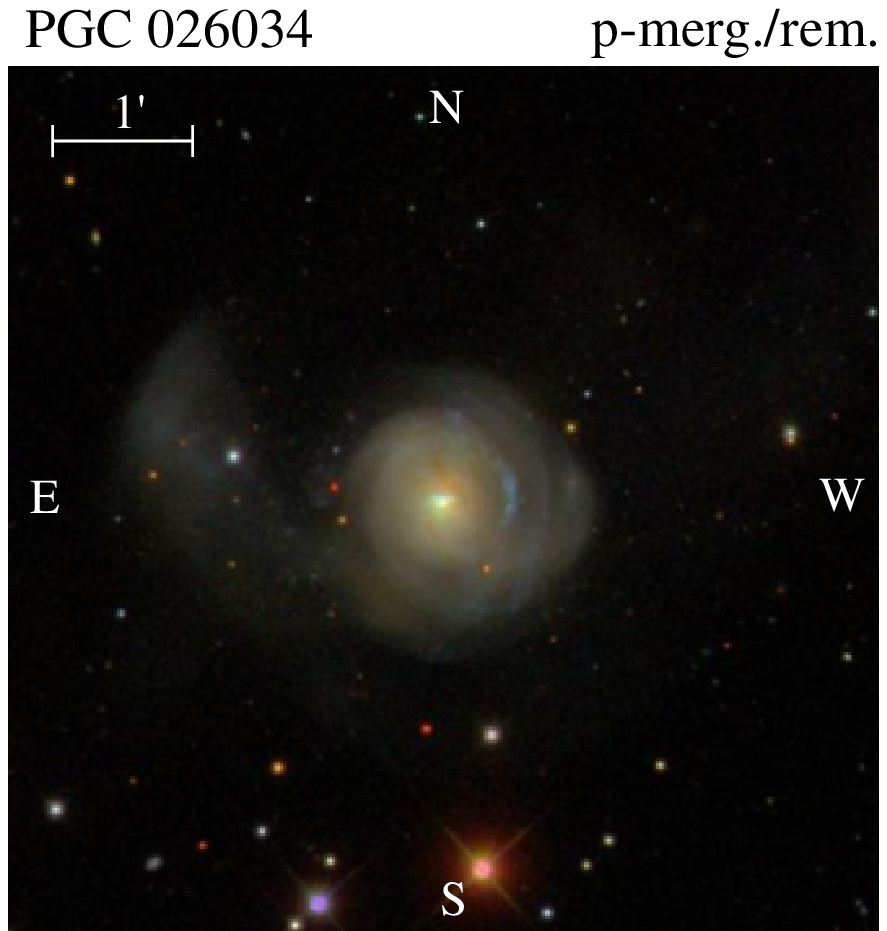}
\end{array}$
\end{center}
\caption{SDSS images representing examples of
         different levels of disturbances of SNe host galaxies.
         The Principal Galaxy Catalogue (PGC) objects' identifiers and
         levels of disturbances are listed at the top.
         In all images, north is up and east to the left.}
\label{disturbexamples}
\end{figure*}

We additionally define for all 608 host galaxies
their level of morphological disturbance with
the possible presence of signs of interactions and mergers.
We do this by visual inspection of
the combined SDSS $g$-, $r$-, and $i$-band images of the hosts,
and comparison with the colour images of the simulations of
equal-mass gas-rich disc mergers in \citet{2008MNRAS.391.1137L}.
The comparison with the simulations is done in order to identify
the stages of interaction for all the host galaxies.
We define four categories of
SN host disturbances:
normal (hosts without any visible disturbance in their
morphological structure), perturbed (hosts with visible
morphological disturbance, but without long tidal arms, bridges,
or destruction of spiral patterns),
interacting (hosts with obvious signs of galaxy--galaxy interaction),
merging (hosts with evidence of an ongoing merging process;
see e.g. \citealt{2012A&A...539A..45L}),
and post-merging/remnant (single galaxies that
exhibit signs of a past interaction,
with a strong or already relaxed disturbance,
see e.g. \citealt{2013MNRAS.435.3627E} and \citealt{2008MNRAS.391.1137L}, respectively).
Fig.~\ref{disturbexamples}
shows images of typical examples of SNe host galaxies
with different levels of
disturbances\footnote{{\footnotesize The full table of
disturbance levels for 608 individual hosts
is only available online.}}.

\begin{table}
  \centering
  \begin{minipage}{78mm}
  \caption{Numbers of SNe in barred
           and unbarred galaxies among the various morphological types.}
  \tabcolsep 2.8pt
  \label{bar-morph}
    \begin{tabular}{lrrrrrrrrrrr}
    \hline
  &\multicolumn{1}{c}{S0/a}&\multicolumn{1}{c}{Sa}
  &\multicolumn{1}{c}{Sab}&\multicolumn{1}{c}{Sb}&\multicolumn{1}{c}{Sbc}&\multicolumn{1}{c}{Sc}&\multicolumn{1}{c}{Scd}
  &\multicolumn{1}{c}{Sd}&\multicolumn{1}{c}{Sdm}&\multicolumn{1}{c}{Sm}&\multicolumn{1}{r}{All}\\
  \hline
    Unbarred & 21 & 19 & 21 & 59 & 104 & 173 & 48 & 15 & 11 & 11 & 482 \\
    Barred & 9 & 6 & 9 & 32 & 45 & 38 & 11 & 40 & 14 & 6 & 210 \\
  \hline
    All & 30 & 25 & 30 & 91 & 149 & 211 & 59 & 55 & 25 & 17 & 692 \\
  \hline \\
  \end{tabular}
  \end{minipage}
\end{table}

The visual inspection of
the combined SDSS $g$-, $r$-, and $i$-band images
is performed by one of the co-authors
(TAN) only, in order to insure a uniform classification.
To test this classification, a subset of 100 host galaxies
is randomly selected from the classified
sample, and re-classified by another co-author (AAH).
By comparing both classifications for the 100 galaxy subset,
we measure a difference of disturbance class by one unit in only 5 per cent
of the galaxies, suggesting that there are no major systematic
biases in the full classification.
Table~\ref{disturb-morph} shows the distribution of
SNe in the 608 host galaxies of different morphological
types among the different disturbance classes.

In addition, using the standard diagnostic diagram
(\citealt*[][hereafter BPT]{1981PASP...93....5B})
presented in \citetalias{2012A&A...544A..81H},
we were able to determine different
activity classes of nuclei of SNe host galaxies.
Inspecting the SDSS DR10 \citep{2014ApJS..211...17A},
we find additional spectra of 12 host galaxies
from the Baryon Oscillation Spectroscopic Survey.
Using the method described in \citetalias{2012A&A...544A..81H},
we also classify 11 of them (see Table~\ref{hosts-new-BPT}).
In the BPT diagram, the hosts with SDSS nuclear
spectra include 46~narrow-line AGN (10~Sy and 36~LINER), 143~SF,
and 43~C galaxies with 268~SNe in total.
The spectra of 31 galaxies with 31 SNe did not meet the criteria on
signal-to-noise ratio of the BPT diagnostic (see \citetalias{2012A&A...544A..81H})
and thus were not analysed.
In total, the activity classes of nuclei of 345 galaxies with 393 SNe
were not determined due to absence or poor quality
(nine spectra have bad flux calibration or bad redshift determination) of their SDSS spectra.
Table~\ref{BPT-morph} displays the distribution of
SNe in hosts with different activity classes of nuclei
among the various morphological types.

\begin{table}
  \centering
  \begin{minipage}{77mm}
  \caption{Numbers of SNe in galaxies with different disturbance levels
           among the different morphological types.}
  \tabcolsep 2.2pt
  \label{disturb-morph}
    \begin{tabular}{lrrrrrrrrrrr}
    \hline
  &\multicolumn{1}{c}{S0/a}&\multicolumn{1}{c}{Sa}
  &\multicolumn{1}{c}{Sab}&\multicolumn{1}{c}{Sb}&\multicolumn{1}{c}{Sbc}&\multicolumn{1}{c}{Sc}&\multicolumn{1}{c}{Scd}
  &\multicolumn{1}{c}{Sd}&\multicolumn{1}{c}{Sdm}&\multicolumn{1}{c}{Sm}&\multicolumn{1}{r}{All}\\
  \hline
    Normal & 22 & 13 & 18 & 63 & 104 & 151 & 35 & 39 & 15 & 10 & 470 \\
    Perturbed & 0 & 5 & 4 & 20 & 33 & 41 & 20 & 15 & 9 & 2 & 149 \\
    Interacting & 3 & 1 & 6 & 3 & 9 & 7 & 3 & 1 & 1 & 4 & 38 \\
    Merging & 0 & 0 & 0 & 2 & 2 & 12 & 1 & 0 & 0 & 0 & 17 \\
    p-merg./rem. & 5 & 6 & 2 & 3 & 1 & 0 & 0 & 0 & 0 & 1 & 18 \\
  \hline
    All & 30 & 25 & 30 & 91 & 149 & 211 & 59 & 55 & 25 & 17 & 692 \\
  \hline \\
  \end{tabular}
  \end{minipage}
\end{table}
\begin{table}
  \centering
  \begin{minipage}{42mm}
  \caption{New activity classes of nuclei for 11 host
           galaxies which were listed without classification
           in \citetalias{2012A&A...544A..81H}.}
    \label{hosts-new-BPT}
    \begin{tabular}{ll}
    \hline
  \multicolumn{1}{c}{Galaxy$^a$}&\multicolumn{1}{c}{BPT}\\
  \hline
    J075126.19+140113.6 & LINER \\
    J075923.61+162516.7 & SF \\
    J094041.64+115318.1 & C \\
    J095619.16+164952.2 & SF \\
    J100157.94+554047.8 & LINER \\
    J103710.21+123909.2 & SF \\
    J110312.95+110436.3 & SF \\
    J120236.52+410315.0 & SF \\
    J134913.75+351526.2 & C \\
    J150821.40+215245.3 & SF \\
    J155452.10+210700.0 & SF \\
  \hline \\
  \end{tabular}
  \parbox{\hsize}{
   $^a$Host galaxy SDSS designation.}
  \end{minipage}
\end{table}

\section{Results and discussion}
\label{Resdiscus}

In this section, we examine the $N_{\rm Ia}/N_{\rm CC}$,
$N_{\rm Ibc}/N_{\rm II}$, and $N_{\rm Ic}/N_{\rm Ib}$
ratios of the different types of SNe as a function of the host galaxy properties
(morphology, presence of bar, disturbance, and activity class of nucleus).
In the second column of Table~\ref{SNratios-morph-table_s}, we show
the calculated number ratios of SN types
for all hosts of types S0/a--Sm.
The errors of number ratios are calculated using
the approach of \citet{2011PASA...28..128C}.

The relative frequencies of SNe types presented in the second column of
Table~\ref{SNratios-morph-table_s} are consistent with previously published values.
The ratio of Type Ia to CC SNe in our sample is $0.44_{-0.03}^{+0.04}$,
which is close to $0.40\pm0.04$ in \citet{2009A&A...503..137B} and $0.40\pm0.08$
obtained from the local SNe sample of \citet{2009MNRAS.395.1409S}.
The $N_{\rm Ibc}/N_{\rm II}$ ratio in our study is $0.36_{-0.03}^{+0.04}$,
which is similar to $0.31\pm0.04$ in \citet{2009A&A...503..137B},
$0.33\pm0.05$ in \citet{2008Ap.....51...69H}, $\sim0.33$ in \citet{2003astro.ph..1006H},
$0.42\pm0.09$ in \citet{2009MNRAS.395.1409S}, $\sim0.35$ in \citet{2011MNRAS.412.1522S},
and also is similar to those obtained from
ratios of CC SN rates in spiral galaxies
\citep[e.g.][]{2005A&A...433..807M,2011Ap.....54..301H,2011MNRAS.412.1473L}.
Finally, our result for $N_{\rm Ic}/N_{\rm Ib}$ ratio is $2.12_{-0.42}^{+0.48}$,
consistent within larger error bars with $1.53\pm0.35$ in \citet{2009A&A...503..137B},
$2.0\pm0.8$ in \citet{2009MNRAS.395.1409S}, and $\sim2.1$ in \citet{2011MNRAS.412.1522S}.

\subsection{Dependence of relative frequencies of SNe types on host morphology}

\begin{table}
  \centering
  \begin{minipage}{78mm}
  \caption{Numbers of SNe in galaxies with
  different activity classes of nuclei among the various morphological types.
  The activity classes are available for only 232 hosts
  among the sample of 608 host galaxies.}
  \tabcolsep 3.2pt
    \label{BPT-morph}
    \begin{tabular}{lrrrrrrrrrrr}
    \hline
  &\multicolumn{1}{c}{S0/a}&\multicolumn{1}{c}{Sa}
  &\multicolumn{1}{c}{Sab}&\multicolumn{1}{c}{Sb}&\multicolumn{1}{c}{Sbc}&\multicolumn{1}{c}{Sc}&\multicolumn{1}{c}{Scd}
  &\multicolumn{1}{c}{Sd}&\multicolumn{1}{c}{Sdm}&\multicolumn{1}{c}{Sm}&\multicolumn{1}{r}{All}\\
  \hline
    SF & 3 & 3 & 4 & 12 & 27 & 42 & 26 & 28 & 11 & 5 & 161 \\
    C & 1 & 2 & 2 & 9 & 9 & 27 & 6 & 0 & 1 & 0 & 57 \\
    Sy & 2 & 0 & 0 & 3 & 3 & 2 & 0 & 0 & 0 & 0 & 10 \\
    LINER & 3 & 2 & 3 & 10 & 12 & 10 & 0 & 0 & 0 & 0 & 40 \\
  \hline
    All & 9 & 7 & 9 & 34 & 51 & 81 & 32 & 28 & 12 & 5 & 268 \\
  \hline \\
  \end{tabular}
  \end{minipage}
\end{table}
\begin{table}
  \centering
  \begin{minipage}{82mm}
  \caption{The number ratios of SN types ($N_{\rm Ia}/N_{\rm CC}$, $N_{\rm Ibc}/N_{\rm II}$,
  and $N_{\rm Ic}/N_{\rm Ib}$) in all spiral galaxies and in two morphological subsamples, and
  the significance value $P_{\rm B}$ of the difference between the ratios in the two subsamples.
  The total number of SNe is 692 in 608 host galaxies.}
  \label{SNratios-morph-table_s}
    \begin{tabular}{lrrrr}
    \hline
  & \multicolumn{1}{c}{All} & \multicolumn{1}{c}{S0/a--Sbc} & \multicolumn{1}{c}{Sc--Sm} & \multicolumn{1}{c}{$P_{\rm B}$} \\
  \hline
    $N_{\rm Ia}/N_{\rm CC}$ & $0.44_{-0.03}^{+0.04}$ & $0.70_{-0.07}^{+0.08}$ & $0.26_{-0.03}^{+0.04}$ & $3 \times 10^{-9}$ \\
    $N_{\rm Ibc}/N_{\rm II}$ & $0.36_{-0.03}^{+0.04}$ & $0.45_{-0.06}^{+0.08}$ & $0.31_{-0.04}^{+0.05}$ & \multicolumn{1}{c}{0.05} \\
    $N_{\rm Ic}/N_{\rm Ib}$ & $2.12_{-0.42}^{+0.48}$ & $2.75_{-0.82}^{+0.98}$ & $1.75_{-0.44}^{+0.53}$ & \multicolumn{1}{c}{0.22} \\
  \hline \\
  \end{tabular}
  \end{minipage}
\end{table}
\begin{figure*}
  \begin{center}$
  \begin{array}{cc}
  \includegraphics[width=0.47\hsize]{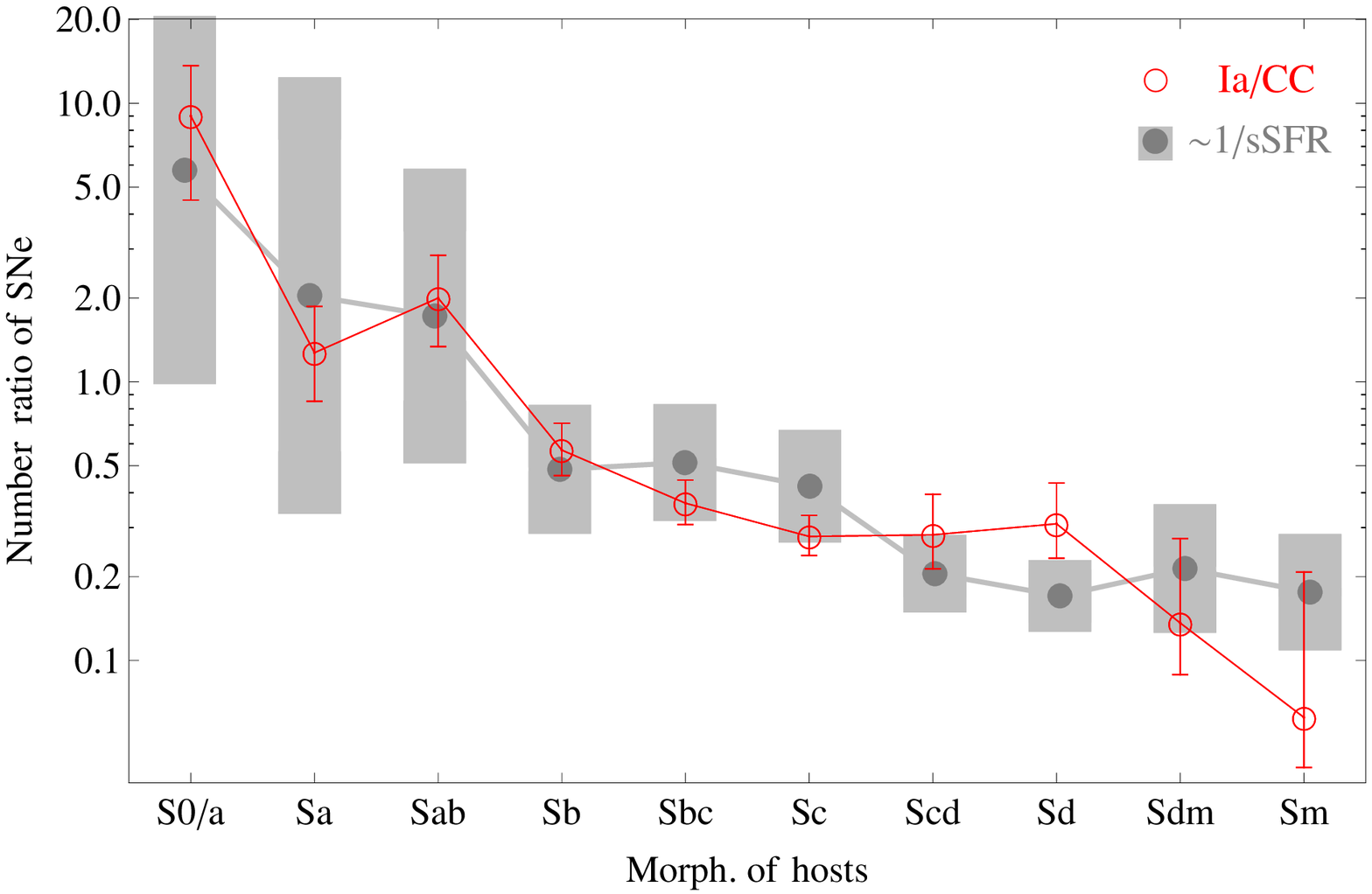} &
  \includegraphics[width=0.47\hsize]{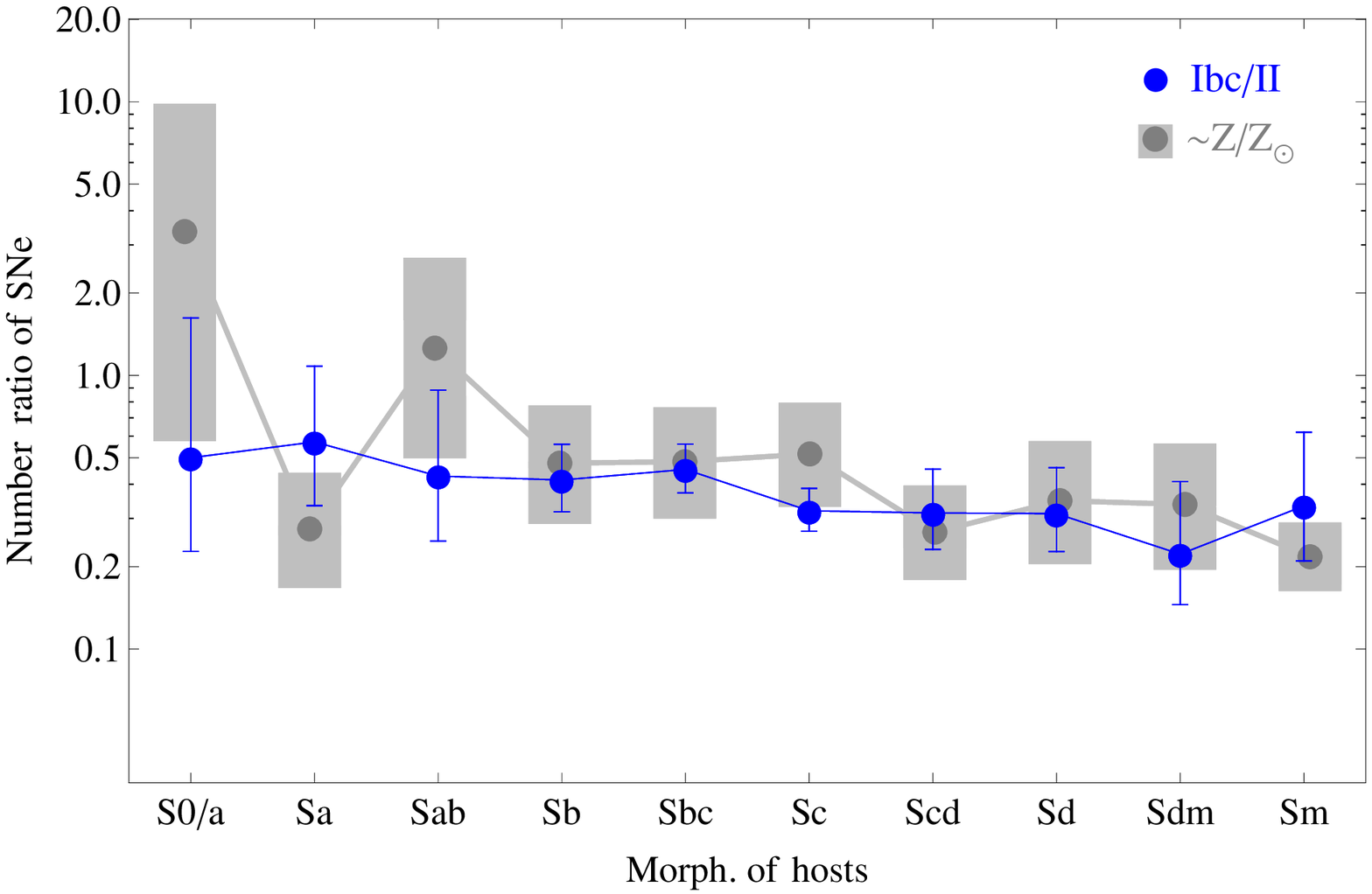}
  \end{array}$
  \end{center}
  \caption{Relative frequency of SNe types as a function of
  host-galaxy morphology. The number ratio of Type Ia to
  CC SNe is presented with red open circles (left-hand panel), while the
  ratio of Types Ibc to II SNe is presented with blue filled circles (right-hand panel).
  The mean values of 1/sSFR (left-hand panel) and ${\rm Z/Z_\odot}$ (right-hand panel) of
  host galaxies are presented
  with dark grey circles. The light grey vertical bars reflect errors of the mean values.
  These distributions are shifted (not scaled) towards the vertical axis to visually
  fit the $N_{\rm Ia}/N_{\rm CC}$ and $N_{\rm Ibc}/N_{\rm II}$ distributions, respectively.}
  \label{ratiottype}
\end{figure*}
\begin{table*}
  \centering
  \begin{minipage}{161mm}
  \caption{The number ratios of SN types ($N_{\rm Ia}/N_{\rm CC}$
           and $N_{\rm Ibc}/N_{\rm II}$) according to
           morphological classification of host galaxies.
           The total number of SNe is 692 in 608 host galaxies.}
  \tabcolsep 3.5pt
  \label{SNratios-morph-table}
    \begin{tabular}{lrrrrrrrrrr}
    \hline
  &\multicolumn{1}{c}{S0/a}&\multicolumn{1}{c}{Sa}
  &\multicolumn{1}{c}{Sab}&\multicolumn{1}{c}{Sb}&\multicolumn{1}{c}{Sbc}&\multicolumn{1}{c}{Sc}&\multicolumn{1}{c}{Scd}
  &\multicolumn{1}{c}{Sd}&\multicolumn{1}{c}{Sdm}&\multicolumn{1}{c}{Sm}\\
  \hline
    $N_{\rm Ia}/N_{\rm CC}$ & $9.00_{-4.51}^{+4.64}$ & $1.27_{-0.42}^{+0.59}$ & $2.00_{-0.66}^{+0.85}$ & $0.57_{-0.11}^{+0.14}$ & $0.37_{-0.06}^{+0.08}$ & $0.28_{-0.04}^{+0.05}$ & $0.28_{-0.07}^{+0.11}$ & $0.31_{-0.08}^{+0.12}$ & $0.14_{-0.05}^{+0.14}$ & $0.06_{-0.02}^{+0.15}$ \\
    $N_{\rm Ibc}/N_{\rm II}$ & $0.50_{-0.27}^{+1.12}$ & $0.57_{-0.24}^{+0.51}$ & $0.43_{-0.18}^{+0.45}$ & $0.41_{-0.10}^{+0.15}$ & $0.45_{-0.08}^{+0.11}$ & $0.32_{-0.05}^{+0.07}$ & $0.31_{-0.08}^{+0.14}$ & $0.31_{-0.09}^{+0.15}$ & $0.22_{-0.08}^{+0.19}$ & $0.33_{-0.12}^{+0.29}$ \\
  \hline \\
  \end{tabular}
  \end{minipage}
\end{table*}

In Table~\ref{SNratios-morph-table},
we present the distribution of
number ratios of SN types according to the morphological
classification of the host galaxies.
Here, we do not calculate the $N_{\rm Ic}/N_{\rm Ib}$ ratio due to
insufficient number of Ib and Ic SNe in each morphological bin
(see also Table~\ref{SN-morph-table}).
The values of $N_{\rm Ia}/N_{\rm CC}$ and $N_{\rm Ibc}/N_{\rm II}$
as a function of morphological type
are shown in Fig.~\ref{ratiottype}.

As can be seen from the left-hand panel of Fig.~\ref{ratiottype},
there is a strong trend in the behaviour of $N_{\rm Ia}/N_{\rm CC}$
depending on host-galaxy morphological types,
such that early-type spirals
include proportionally more Type Ia SNe.
As shown in Table~\ref{SNratios-morph-table_s},
the difference between the number ratios of $N_{\rm Ia}/N_{\rm CC}$
in two broad morphology bins, i.e., early-type (S0/a--Sbc)
and late-type (Sc--Sm) spirals,
is statistically significant.
The significance value $P_{\rm B}$ is calculated using Barnard's exact test\footnote{{\footnotesize
Barnard's test \citep{1945Natur.156..177B} is a non-parametric test for $2 \times 2$
contingency tables, often considered more powerful than Fisher's exact test.}},
which compares the pairs of numbers rather than the number ratios.
A similar behaviour of the relative frequencies of SNe types was found
by \citet{2009A&A...503..137B},
who considered luminosity instead of morphological type.
These authors found that brighter galaxies host
proportionally more Ia SNe than CC SNe.
Their sample consists of 582 SNe in galaxies with
$V_{\rm hel} < 5000~{\rm km~s^{-1}}$ ($\lesssim70~{\rm Mpc}$) and morphological types
corresponding to spirals (S0 to Sd) and irregulars (Irr).

\citet{2005ApJ...629L..85S}
specified that in spiral galaxies, $N_{\rm Ia}$
is a linear combination of the SFR and the total stellar mass ($M_{\ast}$).
In addition, it is well known that $N_{\rm CC}$ is proportional to the SFR
\citep[e.g.][]{2005A&A...433..807M}.
Thus, $N_{\rm Ia}/N_{\rm CC}$ is expected to depend linearly on $M_{\ast}/{\rm SFR}$.
To qualitatively demonstrate the relation between the $N_{\rm Ia}/N_{\rm CC}$ ratio and
the specific SFR (${\rm sSFR}\equiv {\rm SFR}/M_{\ast}$) of host galaxies,
we also display in the left-hand panel of Fig.~\ref{ratiottype}
the distribution of 1/sSFR as a function of morphology, as dark grey circles.
These values are calculated as the mean over morphological types.
The sSFR\footnote{{\footnotesize
The SDSS best-fitting sSFR values were estimated on extinction-corrected \emph{ugriz} model magnitudes
scaled to the \emph{i}-band \emph{c-model} magnitude,
using the publicly available Flexible Stellar Population Synthesis code
(\texttt{FSPS}; \citealt*{2009ApJ...699..486C}).}}
values are extracted for the host galaxies of 253 SNe with available spectra in the SDSS DR10.
The light grey vertical bars in the left-hand panel of Fig.~\ref{ratiottype}
are the errors of the mean values of 1/sSFR for each bin.
The whole distribution is shifted (not scaled) towards the vertical axis to visually
match the $N_{\rm Ia}/N_{\rm CC}$ distribution.
As for $N_{\rm Ia}/N_{\rm CC}$, there is a strong trend in
the distribution of 1/sSFR, such that the sSFR of host galaxies
systematically increases from early- (high-mass or high-luminosity)
to late-type (low-mass or low-luminosity) spirals.
Here, we share the view with \citet{2009A&A...503..137B}
that massive (early-type) spirals have, on average, lower sSFR because of
their smaller gas fractions \citep[e.g.][]{2001MNRAS.321..733B}.
In this picture, the behaviour of $N_{\rm Ia}/N_{\rm CC}$ versus morphology
is a simple reflection of the behaviour of 1/sSFR versus morphological type of galaxies.

The right-hand panel of Fig.~\ref{ratiottype} presents the distribution of
the relative frequency of Types Ibc to II SNe versus host's morphology.
The distribution is nearly flat and shows no apparent dependence on the morphological types.
In linear units, the $N_{\rm Ibc}/N_{\rm II}$ ratio is slightly decreasing
when host morphology becomes of later type.
Therefore, when we divide the host sample into two broad morphology bins,
the difference between the number ratios in these bins becomes
significant (see Table~\ref{SNratios-morph-table_s}).
The trend is similar for $N_{\rm Ic}/N_{\rm Ib}$, such that early-type
spirals include proportionally more Type Ic than Type Ib SNe.
But the difference in the relative frequencies of these SNe types is not significant
(see Table~\ref{SNratios-morph-table_s})
probably due to the small number statistics in the Types Ib and Ic SNe.
Thus, in comparison with the $N_{\rm Ia}/N_{\rm CC}$ ratio,
the $N_{\rm Ibc}/N_{\rm II}$
ratio shows weaker variation with morphology of the hosts.

In general, the $N_{\rm Ibc}/N_{\rm II}$ ratio depends on metallicity, age,
the fraction of binary systems etc.
(e.g. \citealt{2002MNRAS.331L..25B};
\citealt*{2008MNRAS.384.1109E};
\citealt*{2011MNRAS.414.3501E};
\citealt{2011MNRAS.412.1522S}).
To qualitatively examine the relationship between this ratio and metallicity,
we use the extracted global galaxy
metallicities\footnote{{\footnotesize
The method of best-fitting global galaxy
metallicity (Z) estimation is based on
the SDSS photometric and spectroscopic data, and
uses the \texttt{FSPS} code by \citet[][]{2009ApJ...699..486C}.}} available
for only 196 CC SNe hosts from the SDSS DR10.
In the right-hand panel of Fig.~\ref{ratiottype},
the mean values of global metallicity and their uncertainties are represented
by dark grey circles and light grey vertical bars, respectively.
Again, the whole distribution is shifted (not scaled) towards the vertical axis
to ease visual comparison.
The large scatter of metallicity in the first three bins of morphology is attributed
to the small numbers statistics of CC SNe host galaxies with the available SDSS data.
As can be seen, even for our smaller subsample,
the metallicity decreases not significantly from early- to late-type hosts, which reflects
the observed mass--morphology \citep[e.g.][]{2010ApJS..186..427N} and
mass--metallicity \citep[e.g.][]{2004ApJ...613..898T} relations
for spiral galaxies, predicting $\lesssim 0.5$~dex variations.
Therefore, one may interpret the behaviour of $N_{\rm Ibc}/N_{\rm II}$
ratio versus morphology as a possible reflection of
the behaviour of global metallicity versus morphological type of spirals.
However, we emphasize that any metallicity constraint from our analyses
is very weak.

\begin{table}
  \centering
  \begin{minipage}{66mm}
  \caption{The number ratios of SN types
           in low- and high-luminosity bins of hosts,
           and the significance value $P_{\rm B}$ of the difference between the ratios in the two subsamples.
           The total number of SNe is 692 in 608 host galaxies.}
  \label{SNratios-Mg-table}
    \begin{tabular}{lrrr}
    \hline
  & \multicolumn{1}{c}{$M_{\rm g}>-20.7$} & \multicolumn{1}{c}{$M_{\rm g}\leqslant-20.7$} & \multicolumn{1}{c}{$P_{\rm B}$} \\
  \hline
    $N_{\rm Ia}/N_{\rm CC}$ & $0.39_{-0.04}^{+0.05}$ & $0.48_{-0.05}^{+0.06}$ & 0.12 \\
    $N_{\rm Ibc}/N_{\rm II}$ & $0.34_{-0.04}^{+0.05}$ & $0.39_{-0.05}^{+0.06}$ & 0.28 \\
    $N_{\rm Ic}/N_{\rm Ib}$ & $1.76_{-0.47}^{+0.58}$ & $2.53_{-0.69}^{+0.82}$ & 0.29 \\
  \hline \\
  \end{tabular}
  \end{minipage}
\end{table}

The dependence of the SNe number ratios
on the global metallicity of the host galaxies
previously was studied
(e.g. \citealt{2003A&A...406..259P};
\citealt*{2008ApJ...673..999P};
\citealt{2009A&A...503..137B,2012ApJ...759..107K}).
\citet{2009A&A...503..137B} derived global galaxy
gas-phase metallicities (oxygen abundances),
using the well-known metallicity--luminosity
relation, and confirmed their earlier finding \citep{2003A&A...406..259P}, namely
that there is a positive correlation between $N_{\rm Ibc}/N_{\rm II}$ and metallicity.
The size of their sample did not allow firm conclusions
about the $N_{\rm Ic}/N_{\rm Ib}$ ratio.
They explained these results using the following interpretation:
with increasing galaxy metallicity (luminosity), the stellar envelope is
lost more easily and lower mass stars may become SNe of Type Ibc (Ic),
thus increasing the $N_{\rm Ibc}/N_{\rm II}$ ($N_{\rm Ic}/N_{\rm Ib}$) ratio.
Similar results were obtained for the $N_{\rm Ibc}/N_{\rm II}$
versus metallicity relation by \citet{2008ApJ...673..999P},
who did not use the metallicity--luminosity relation but the directly measured
metallicity for the hosts of 115 SNe
from the SDSS DR4 in the redshift range of $0.01 < z < 0.04$.
However, using spectra of 74 host H~{\footnotesize II} regions,
\citet{2010MNRAS.407.2660A} found that the mean metallicity of
immediate environments (which can be assumed to be more accurate)
of SNe Ibc is not significantly higher ($\sim0.06$~dex)
than that of SNe II. They suggested a possible metallicity sequence,
in terms of increasing progenitor metallicity going from SNe II through
SNe Ib and finally SNe Ic.
Based on the sample of 20 well-observed SNe Ib/c, \citet{2011A&A...530A..95L}
detected no significant difference between metallicities of immediate
environments of Types Ic (higher in average by 0.08~dex) and Ib SNe.
\citet{2011ApJ...731L...4M} studied 35 SNe Ib/c obtaining significantly higher
environmental metallicities for Type Ic SNe. However,
\citet{2011A&A...530A..95L} noted that the results of \citet{2011ApJ...731L...4M}
were affected by the uncertainties in the individual metallicity measurements.
\citet{2012ApJ...758..132S} also found an insignificant difference between
median metallicities from the spectra of immediate environments of
12 SNe Ib and 21 SNe Ic.

In addition, \citet{2009A&A...503..137B} showed that the
$N_{\rm Ia}/N_{\rm CC}$ ratio increases with the metallicity of the host galaxies.
However, they noted that the metallicities of SNe hosts simply
reflect the observed mass--metallicity relation \citep[e.g.][]{2004ApJ...613..898T},
and does not directly affect the $N_{\rm Ia}/N_{\rm CC}$ ratio
in contrast to the case of $N_{\rm Ibc}/N_{\rm II}$ ratio.
They suggest that it is the mass of hosts that affects the $N_{\rm Ia}/N_{\rm CC}$ ratio.

In the above-mentioned studies, luminosities or metallicities of the galaxies
were not defined for an individual morphological bin.
Therefore, each luminosity or metallicity bin may have wide scatter of morphological
types of the host galaxies.
In this respect, to directly check the correlation of the number ratios of
the different SN types with luminosity, one can also calculate
the number ratios in two bins of low and high luminosities, independently of the morphology.
Note that the luminosities are available for all the host galaxies,
while the masses, for only about one-third of the sample.
The SDSS $g$-band absolute magnitudes of galaxies
were calculated using the data from \citetalias{2012A&A...544A..81H}.
Table~\ref{SNratios-Mg-table} shows the relative frequencies of SNe types in
low- and high-luminosity bins.
Sizes of the bins are constructed to include nearly equal numbers of all types of SNe in each bin.
A similar analysis using the extracted metallicities of the host galaxies
was not meaningful due to the smaller subsample
(only 270 SNe have the host galaxies with a measured metallicity in the SDSS),
as well as lower precision of the metallicity measurements compared to
the luminosity measurements of the SNe hosts.

\begin{figure}
  \begin{center}
  \includegraphics[width=0.95\hsize]{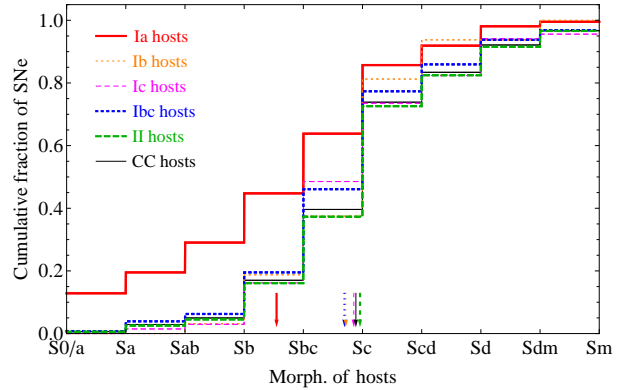}
  \end{center}
  \caption{Cumulative fractions of the different types of SNe (692 events) versus
           host galaxy morphological type. The mean morphologies of
           host galaxies of each SNe type are shown by arrows.}
  \label{SNehostcum}
\end{figure}
\begin{table}
  \centering
  \begin{minipage}{75mm}
  \caption{KS test probabilities of consistency
           for the distributions of hosts morphologies
           among different subsamples of SN types.
           The mean values of the morphological \emph{t}-type
           and their errors are presented in parentheses.}
  \label{SNehosts_P_table}
    \begin{tabular}{l@{\hspace{2mm}}lll@{\hspace{2mm}}ll}
    \hline
  \multicolumn{2}{c}{Subsample~1} & & \multicolumn{2}{c}{Subsample~2} & \multicolumn{1}{c}{$P_{\rm KS}$} \\
  \hline
    Ia & ($3.5\pm0.1$) & versus & Ib & ($4.7\pm0.2$) & \multicolumn{1}{c}{0.04} \\
    Ia & ($3.5\pm0.1$) & versus & Ic & ($4.9\pm0.2$) & $5 \times 10^{-5}$ \\
    Ia & ($3.5\pm0.1$) & versus & Ibc & ($4.7\pm0.1$) & $8 \times 10^{-6}$ \\
    Ia & ($3.5\pm0.1$) & versus & II & ($5.0\pm0.1$) & $8 \times 10^{-10}$ \\
    Ia & ($3.5\pm0.1$) & versus & CC & ($4.9\pm0.1$) & $3 \times 10^{-10}$ \\
    Ib & ($4.7\pm0.2$) & versus & Ic & ($4.9\pm0.2$) & \multicolumn{1}{c}{0.94} \\
    Ib & ($4.7\pm0.2$) & versus & II & ($5.0\pm0.1$) & \multicolumn{1}{c}{0.85} \\
    Ic & ($4.9\pm0.2$) & versus & II & ($5.0\pm0.1$) & \multicolumn{1}{c}{0.47} \\
    Ibc & ($4.7\pm0.1$) & versus & II & ($5.0\pm0.1$) & \multicolumn{1}{c}{0.46} \\
  \hline \\
  \end{tabular}
  \end{minipage}
\end{table}

In Table~\ref{SNratios-Mg-table}, the $N_{\rm Ia}/N_{\rm CC}$ ratio appears to be
higher for brighter, i.e., high-mass, galaxies.
The result is in agreement with that of \citet{2009A&A...503..137B}.
However, in both cases, the differences are not significant.
The CC SNe are also equally distributed in these two bins.
The $N_{\rm Ibc}/N_{\rm II}$ and $N_{\rm Ic}/N_{\rm Ib}$ ratios are
not significantly higher in brighter galaxies. These
results are similar to relative frequencies of CC SNe types versus $M_{\rm B}$ relations
obtained by \citet{2009A&A...503..137B} and \citet{2003A&A...406..259P}.

Despite the degree of subjectivity involved in the morphological classifications,
the relations between the number ratios and morphology are probably tighter than
those between the number ratios and luminosity or metallicity
(see the significance values in
Tables~\ref{SNratios-morph-table_s} and \ref{SNratios-Mg-table}).
Compared with the indirectly estimated metallicity,
the morphological classification performed in \citetalias{2012A&A...544A..81H}
using the combined $g$-, $r$-, and $i$-band images of the SDSS,
i.e., considering to some extent also
the colours of galaxies, is more closely related to the stellar population.
It is also important to note that the `coloured' morphological type of a galaxy
is an observed parameter and is available for the nearby galaxies
thanks to the SDSS imaging, while the sSFR and metallicity of a galaxy are
inferred parameters based on synthetic models of the integrated
broad-band fluxes or the spectra, and are
affected by uncertainties due to dust contamination.
Therefore, we choose to further consider
mainly the observed quantities of the SNe host galaxies.
In particular, we consider the morphology as the most important
parameter shaping the number ratios of the different types of SNe.

Fig.~\ref{SNehostcum} presents the cumulative distributions of
the different SNe versus morphology of the host galaxies.
In particular, 64 per cent of Type Ia SNe were discovered in S0/a--Sbc
galaxies in contrast to 40 per cent of CC SNe.
The mean morphological type of Ia SNe host galaxies is earlier than
those of all types of CC SNe hosts
(see \emph{t}-types in Table~\ref{SNehosts_P_table}
and the corresponding arrows in Fig.~\ref{SNehostcum}).
A KS test shows that the distribution of morphologies for the hosts of Type Ia SNe
is significantly different from those of hosts of all types of CC SNe
(see Table~\ref{SNehosts_P_table}).
In contrast, the mean values and distributions of host morphologies for Type
Ib, Ic, Ibc, and II SNe are nearly identical.
Thus, Type Ia SNe, in contrast to CC SNe,  are
found more frequently in galaxies with older stellar populations.

\subsection{Dependence of relative frequencies of SNe types on host barred structure}
\label{SNeindarredhosts}

Given that bars in galactic stellar discs have an important influence on
the star formation process of galaxies \citep[e.g.][]{2011MNRAS.416.2182E},
we study in this subsection the dependence
of the relative frequencies of SNe types on the presence of bars in spiral host galaxies.
Table~\ref{SNratios-bar-table} shows the distribution of
the number ratios of SN types after separating the hosts into
unbarred and barred types.
There is no significant difference between the $N_{\rm Ia}/N_{\rm CC}$
ratios in the two subsamples. However, the $N_{\rm Ibc}/N_{\rm II}$ ratio is higher,
with barely significance, in unbarred hosts compared with the ratio
in barred hosts. The $N_{\rm Ic}/N_{\rm Ib}$ ratio is higher,
but not significantly, in barred hosts compared with that
in unbarred ones.

\begin{figure}
  \begin{center}
  \includegraphics[width=0.95\hsize]{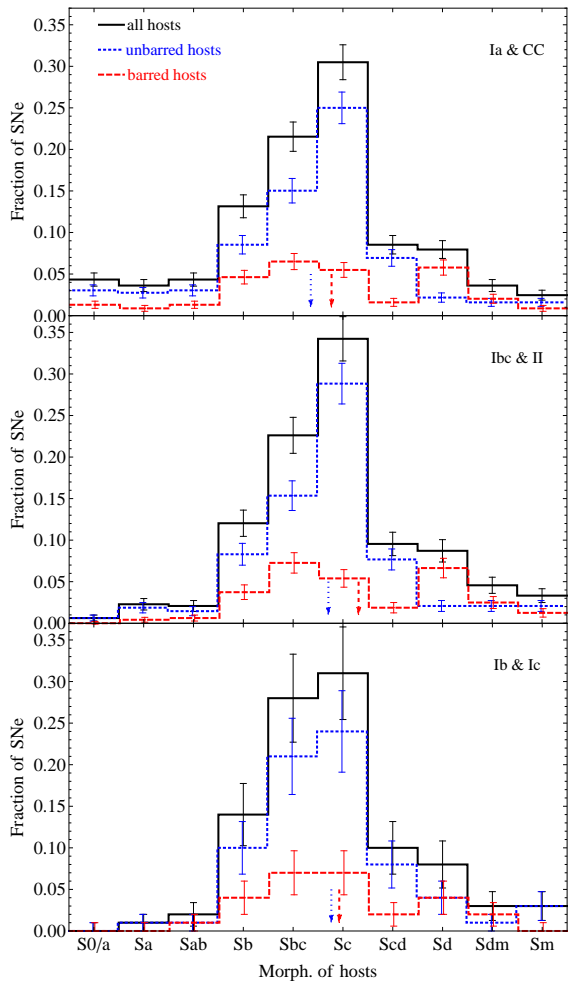}
  \end{center}
  \caption{Fraction of SNe in galaxies with (red dashed lines) and
           without (blue dotted lines) bars, and their sum (black solid lines) versus host galaxy morphological type.
           The distributions of all (Ia and CC), CC (Ibc and II), and Ib plus Ic SNe are presented in the upper,
           middle, and bottom panels, respectively.
           The error bars assume a Poisson distribution
           (with $\pm1$ object if none is found).
           The mean morphologies of host
           galaxies with and without bars are shown by arrows.}
  \label{SNehostBarfract}
\end{figure}

To explain the behaviours of the number ratios of the different SN types
depending on the existence of bar, we analyse the morphological distributions of
host galaxies with and without bars.
Fig.~\ref{SNehostBarfract} shows the distribution of
the various types of SNe in hosts with or without bars
for the different morphological types.
The histograms illustrate that the distribution of SNe in barred galaxies
is bimodal with respect to host-galaxy morphology, with peaks near Sbc and
Sd types (see also Table~\ref{bar-morph}).
In contrast, the distribution of SNe in unbarred galaxies has
a single peak towards Sc type.
A KS test shows that the distributions of morphologies of host galaxies of all types of SNe
with and without bars are significantly different ($P_{\rm KS}=6 \times 10^{-6}$)
despite similar mean values for both distributions
(see arrows in the top panel of Fig.~\ref{SNehostBarfract}).
A similar distribution of SNe host galaxies was already seen
in fig.~8 of \citetalias{2012A&A...544A..81H}.

Moreover, a bimodal distribution of barred galaxies was also found by
\citet{2010ApJ...714L.260N}, who recently released a morphological
catalogue of 14034 visually classified galaxies ($0.01 < z < 0.1$)
from the SDSS \citep{2010ApJS..186..427N}.
They noted that the bar fraction is bimodal with disc galaxy colour,
having peaks towards the redder and bluer galaxies.
The authors suggested that this trend may reveal
two distinct types of bars: strong bars, that are more common in early-type,
massive, redder, and gas-poor discs, and weak bars, that are frequently
found in late-type, low-mass, bluer, and gas-rich spirals.
Also based on the SDSS, but using instead
the length of bars measured in nearby galaxies from the EFIGI\footnote{{\footnotesize
Extraction de Formes Id\'{e}alis\'{e}es de Galaxies en Imagerie.}}
catalogue \citep{2011A&A...532A..74B},
a multiwavelength data base of visually classified 4458 PGC galaxies,
\citet{2011A&A...532A..75D}
found that the fraction of barred galaxies
is bimodal with a strong peak for type Sab, and a
weaker one for type Sd (see their fig.~9).

\begin{table}
  \centering
  \begin{minipage}{58mm}
  \caption{The number ratios of SN types in unbarred
           and barred hosts, and the significance value $P_{\rm B}$ of the difference between
           the ratios in the two subsamples.}
  \label{SNratios-bar-table}
    \begin{tabular}{lrrr}
    \hline
  & \multicolumn{1}{c}{Unbarred} & \multicolumn{1}{c}{Barred} & \multicolumn{1}{c}{$P_{\rm B}$} \\
  \hline
    $N_{\rm Ia}/N_{\rm CC}$ & $0.42_{-0.04}^{+0.05}$ & $0.47_{-0.06}^{+0.08}$ & 0.31 \\
    $N_{\rm Ibc}/N_{\rm II}$ & $0.40_{-0.04}^{+0.05}$ & $0.28_{-0.05}^{+0.07}$ & 0.07 \\
    $N_{\rm Ic}/N_{\rm Ib}$ & $1.92_{-0.43}^{+0.51}$ & $2.86_{-1.07}^{+1.31}$ & 0.29 \\
  \hline \\
  \end{tabular}
  \end{minipage}
\end{table}

To check the consistency of bar detection in our sample with that of the EFIGI catalogue,
we make a subsample of galaxies (231 objects) that are common to both EFIGI and
to our sample of SNe hosts. We compare the EFIGI \texttt{Bar Length} attribute
with our detection (bar or no bar). When the \texttt{Bar Length} is 0, 2, 3, or 4,
our bar detection is different for only 4 per cent of cases.
However, we do not detect bars in 62 per cent of cases when \texttt{Bar Length} is 1.
The EFIGI \texttt{Bar Length} $=$ 1 mainly corresponds to the threshold of our bar detection.
Moreover, to compare the bar fractions in the SNe hosts and the EFIGI catalogue within
each morphological type, we limit the galaxies in both samples to the same distances and exclude
the edge-on ($80^{\circ} \leqslant i \leqslant 90^{\circ}$) galaxies and those with completely distorted profiles
(mergers and post-mergers/remnants) from both samples.
The bar fractions per morphological type are in agreement with those of
the EFIGI within our larger errors and
taking into account that the detection of bar for
\texttt{Bar Length} $=$ 1 is somewhat different.

Thus, the unbarred and barred spirals have `specific'
distributions with morphology, consequently shaping
the SNe distributions in these galaxies.

In particular, the $N_{\rm Ia}/N_{\rm CC}$ ratios in galaxies with and without bars
are nearly equal, because the reduced ratio in barred galaxies from
the second Sd--Sdm peak is somewhat compensated by
the increased ratio in S0/a--Sc
barred hosts (see the upper panel of
Fig.~\ref{SNehostBarfract} and Table~\ref{SNratios-morph-table}).
The situation is different for the $N_{\rm Ibc}/N_{\rm II}$ ratio
because the reduced ratio in the second peak of
barred hosts is not compensated, due to a deficiency of CC SNe in
S0/a--Sab hosts (see the middle panel of
Fig.~\ref{SNehostBarfract} and Table~\ref{SNratios-morph-table}).
The $N_{\rm Ic}/N_{\rm Ib}$ ratio is somewhat higher, but not significantly,
in subsample of barred galaxies than in unbarred ones,
because of relatively weak contribution of SNe Ib
from the barred spirals and particularly from the second peak
(see the bottom panel of Fig.~\ref{SNehostBarfract}
and Table~\ref{SN-morph-table}).
In barred galaxies, we have five Ib SNe only in Sbc--Sc hosts in comparison with
two Ib SNe only in Sdm hosts.
The lack of Ib SNe is attributed to
the small numbers statistics of these events.

To exclude the dependence of the number ratios on morphology,
we analyse the ratios in the most populated
morphological bins, i.e., Sbc$+$Sc.
We find that there is no significant difference in the
various number ratios between unbarred
($N_{\rm Ia}/N_{\rm CC}=0.30_{-0.04}^{+0.05}$ and $N_{\rm Ibc}/N_{\rm II}=0.37_{-0.05}^{+0.06}$)
and barred
($N_{\rm Ia}/N_{\rm CC}=0.36_{-0.07}^{+0.11}$ and $N_{\rm Ibc}/N_{\rm II}=0.36_{-0.08}^{+0.13}$)
Sbc--Sc hosts,
and the values of the number ratios are the same as those in
Table~\ref{SNratios-morph-table}.
Thus, in the whole sequence of spiral galaxies,
the variation of the number ratios of SNe with
the presence of a bar is due to the
bimodality of the morphology distribution in barred galaxies.

\subsection{Dependence of relative frequencies of SNe types on host disturbance}

\begin{figure}
  \begin{center}
  \includegraphics[width=0.95\hsize]{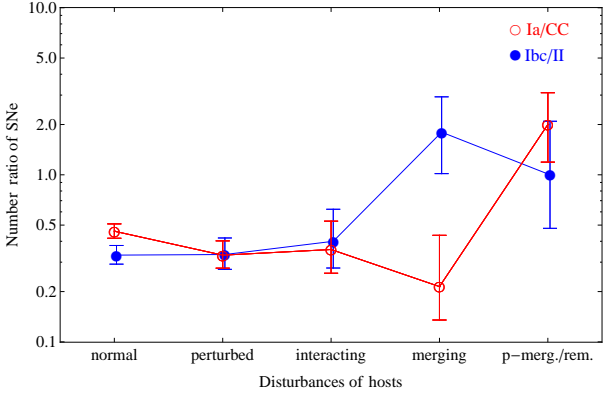}
  \end{center}
  \caption{Relative frequency of SNe types as a function of the
           disturbance level of host galaxies.}
  \label{ratiodisturb}
\end{figure}

Galaxy--galaxy interactions are expected to be responsible
for triggering massive star formation,
especially in strongly disturbed galaxies
\citep[e.g.][]{2005Natur.433..604D,2008MNRAS.391.1137L}.
In this subsection, we study the impact of galaxy host interactions
on the number ratio of the various SNe types,
by using the morphological disturbances of the
hosts, defined in Section~\ref{sample} as normal, perturbed, interacting,
merging and post-merging/remnant, and illustrated in Fig.~\ref{disturbexamples}.
These levels of disturbance are arranged in an approximate
chronological order, according to the different stages of interaction,
and have their own time-scales and levels of star formation
\citep[e.g.][]{2008MNRAS.391.1137L}.
Table~\ref{SNratios-disturbance-table} presents the
values of the $N_{\rm Ia}/N_{\rm CC}$ and $N_{\rm Ibc}/N_{\rm II}$
ratios as a function of disturbance level, and
Fig.~\ref{ratiodisturb} is the corresponding graph.
We do not calculate the $N_{\rm Ic}/N_{\rm Ib}$ ratio due to
insufficient number of Ib and Ic SNe per bin of disturbance.

From Table~\ref{SNratios-disturbance-table} and Fig.~\ref{ratiodisturb},
the $N_{\rm Ia}/N_{\rm CC}$ ratio is at a nearly constant level
when moving from the normal, perturbed to the interacting galaxies.
Then it slightly declines in the merging galaxies and jumps to the highest
value in the post-merging/remnant host galaxies.
Only the difference between the $N_{\rm Ia}/N_{\rm CC}$ ratios in
the merging and post-merging/remnant hosts is statistically significant
($P_{\rm B} = 2 \times 10^{-3}$).
Within 10 hosts in the merging subsample, there are only three Ia and 14 CC SNe,
and in 17 post-merging/remnant galaxies, 12 Ia and six CC SNe.

The most likely reason for relative overpopulation of CC SNe in
merging spiral galaxies is that these galaxies are dominated by
violent starbursts \citep[e.g.][]{2012A&A...539A..45L},
particularly in their circumnuclear
regions \citep[e.g.][]{2001ApJ...559..147S}.
During the relatively short time-scale of the merging stage
($\sim0.5$~Gyr), as predicted by numerical simulations
\citep[e.g.][]{2005Natur.433..604D,2008MNRAS.391.1137L},
the spiral, gas-rich galaxies do not have enough time to produce many Type Ia SNe
(lifetime $>0.5$~Gyr), but can intensively produce CC SNe, assuming short lifetimes
for the CC SNe progenitors (lifetime $<0.1$~Gyr).
Moreover, the positions of CC SNe in our merging hosts mostly coincide
with circumnuclear regions and only in few cases with bright H~{\footnotesize II} regions,
which is in agreement with the previously found central excess of CC SNe in
extremely disturbed or merging galaxies
\citep[e.g.][]{2010ApJ...717..342H,2012MNRAS.424.2841H,
2011MNRAS.416..567A,2012A&A...540L...5H}.

In our sample, the observed number of post-merging/remnant hosts is higher than
the observed number of merging hosts most probably due to the longer time-scale of
the latter stage of interaction
\citep[$\gtrsim1$~Gyr; e.g.][]{2005Natur.433..604D,2008MNRAS.391.1137L}.
The post-merging/remnant galaxies
have enough time to produce many Type Ia SNe, but not enough high-mass
star formation to produce sufficient number of CC SNe.
In the simulations of galaxy--galaxy interactions,
the majority of post-merging galaxies, and especially
the merger remnants, are observed after
the final coalescence with a significant loss of gas content
\citep[e.g.][]{2005Natur.433..604D,2008MNRAS.391.1137L}.
Their SFRs are consistent with the SFRs of early-type spirals
\citep[e.g.][]{2008MNRAS.391.1137L},
in which we observe more Type Ia than CC SNe
(see Table~\ref{SN-morph-table}).
It is important to note that some of the merger remnant galaxies
in very late stages of interaction can be classified as normal
early-type galaxies because of visual similarities
\citep[see also figs.~1 and 2 of][]{2008MNRAS.391.1137L}.
For instance, in our sample of normal galaxies, there are
seven hosts of S0/a--Sab types that are also flagged as possible
merger remnants in late stages.
Remarkably, there are eight SNe in these galaxies, all of Type Ia.
Therefore, the true $N_{\rm Ia}/N_{\rm CC}$ ratio could be even higher
than measured here for the post-merging/remnant subsample.

\begin{table}
  \centering
  \begin{minipage}{80mm}
  \caption{The number ratios of SN types according to the
           disturbance level of host galaxies.
           The total number of SNe is 692 in 608 host galaxies.}
  \tabcolsep 1pt
  \label{SNratios-disturbance-table}
    \begin{tabular}{lrrrrr}
    \hline
  & \multicolumn{1}{c}{normal} & \multicolumn{1}{c}{perturbed} & \multicolumn{1}{c}{interacting} & \multicolumn{1}{c}{merging} &\multicolumn{1}{c}{p-merg./rem.} \\
  \hline
    $N_{\rm Ia}/N_{\rm CC}$ & $0.46_{-0.04}^{+0.05}$ & $0.33_{-0.05}^{+0.07}$ & $0.36_{-0.10}^{+0.17}$ & $0.21_{-0.08}^{+0.22}$ & $2.00_{-0.81}^{+1.10}$ \\
    $N_{\rm Ibc}/N_{\rm II}$ & $0.33_{-0.04}^{+0.05}$ & $0.33_{-0.06}^{+0.09}$ & $0.40_{-0.12}^{+0.22}$ & $1.80_{-0.78}^{+1.13}$ & $1.00_{-0.52}^{+1.09}$ \\
  \hline \\
  \end{tabular}
  \end{minipage}
\end{table}

The $N_{\rm Ibc}/N_{\rm II}$ ratio is nearly constant when moving
from the normal, and perturbed to the interacting hosts.
Then, in contrast to the $N_{\rm Ia}/N_{\rm CC}$ ratio,
it jumps to its highest value in the merging galaxies and
subsequently slightly declines in the post-merging/remnant subsample.
Within seven hosts of CC SNe in the merging subsample,
there are nine SNe Ibc and five SNe II,
and in six post-merging/remnant hosts of CC SNe,
only three SNe Ibc and three SNe II.
Only the difference of the $N_{\rm Ibc}/N_{\rm II}$ ratios between
the interacting and merging hosts is statistically significant ($P_{\rm B} = 0.02$).

Despite the small numbers statistics, we speculate that the remarkable
excess of SNe Ibc within the central regions of strongly disturbed or merging
galaxies \citep[e.g.][]{2010ApJ...717..342H,2012MNRAS.424.2841H,2011MNRAS.416..567A,
2013Ap&SS.347..365N} is responsible for the observed high value of the $N_{\rm Ibc}/N_{\rm II}$ ratio
in the merging subsample.
We find that the merging galaxies hosting CC SNe are
more luminous than the normal ones
(a KS test probability is only 0.03 that they are drawn from the same
distribution) by almost 0.4~mag in the mean of $M_{\rm g}$.
One may interpret this as a possible implication for somewhat
higher metallicity in these galaxies
($\sim0.05$~dex from \citealt{2004ApJ...613..898T}).
However, in the central regions of merging galaxies where the excess of Ibc SNe
is observed, metallicities are systematically lower
($\sim0.2$~dex from \citealt{2006AJ....131.2004K})
than those of normal and weakly interacting galaxies.
The possible roles of metallicity, age, fraction of binary systems, and IMF shape
are discussed in detail in \citet{2011MNRAS.416..567A,2012MNRAS.424.1372A},
\citet{2012MNRAS.424.2841H}, and \citet{2013MNRAS.436.3464K}.
In the final stage of interactions, i.e. in  the post-merging/remnant galaxies,
the observed trend of the $N_{\rm Ibc}/N_{\rm II}$ ratio
probably is caused by the continuous decrease of both gas fraction
and high-mass star formation
\citep[e.g.][]{2005Natur.433..604D,2008MNRAS.391.1137L}.

\begin{figure}
  \begin{center}
  \includegraphics[width=0.95\hsize]{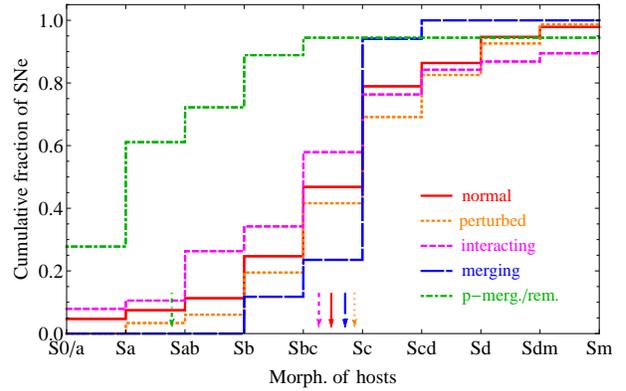}
  \end{center}
  \caption{Cumulative fraction of 692 SNe versus
           host galaxy morphological type. The mean morphologies of
           host galaxies of each type of disturbance are shown by arrows.}
  \label{hostdisturbcum}
\end{figure}

We investigate also how the interaction of galaxies affects
the morphological type of host galaxies.
Fig.~\ref{hostdisturbcum} presents the cumulative distributions of
morphologies of the hosts of 692 SNe for the different disturbance levels.
The mean morphological type of post-merging/remnant
hosts is considerably earlier than
those of all mean types of hosts in different stages of interaction
(see \emph{t}-types in Table~\ref{hosts_disturb_P_table}
and arrows in Fig.~\ref{hostdisturbcum}).
A KS test shows that the distribution of morphology of post-merging/remnant hosts
is significantly different from those of the preceding (merging)
and following (normal) stages (see Table~\ref{hosts_disturb_P_table}).
The mean values and distributions of morphological types of
normal, perturbed, interacting, and merging hosts are not
significantly different from one another.
Thus, the mean morphological classification of the host galaxies is
strongly affected in the final stage of galaxy--galaxy interaction,
respectively, affecting also the number ratios of SNe types.

\begin{table}
  \centering
  \begin{minipage}{81mm}
  \caption{KS test probabilities of consistency
           for the distributions of hosts morphologies
           among the different subsamples by disturbance level.
           The \emph{t}-type mean values and errors of
           the means are presented in parentheses.}
  \tabcolsep 1.8pt
  \label{hosts_disturb_P_table}
    \begin{tabular}{l@{\hspace{1mm}}cll@{\hspace{1mm}}cl}
    \hline
  \multicolumn{2}{c}{Subsample~1} & & \multicolumn{2}{c}{Subsample~2} & \multicolumn{1}{c}{$P_{\rm KS}$} \\
  \hline
    Normal & ($4.5\pm0.1$) & versus & Perturbed & ($4.9\pm0.1$) & \multicolumn{1}{c}{0.23} \\
    Perturbed & ($4.9\pm0.1$) & versus & Interacting & ($4.3\pm0.4$) & \multicolumn{1}{c}{0.17} \\
    Interacting & ($4.3\pm0.4$) & versus & Merging & ($4.7\pm0.2$) & \multicolumn{1}{c}{0.12} \\
    Merging & ($4.7\pm0.2$) & versus & p-merg./rem. & ($1.8\pm0.5$) & $6 \times 10^{-6}$ \\
    p-merg./rem. & ($1.8\pm0.5$) & versus & Normal & ($4.5\pm0.1$) & \multicolumn{1}{c}{$10^{-6}$} \\
  \hline \\
  \end{tabular}
  \end{minipage}
\end{table}

A similar behaviour of morphology is predicted in the simulations of
interactions: the initial galaxies with late-type disc morphologies
become perturbed during the first passage, reach both the maximum morphological
disturbances and SFR at the final merger, then the post-merger and remnant
galaxies gradually end up with early-type disc morphologies and gas content
\citep[e.g.][]{2005Natur.433..604D,2008MNRAS.391.1137L}.
We understand that although the duration and strength of the observed morphological
disturbances and SFRs depend on the mass, merger orientation, and orbital parameters,
the gas properties of the initial galaxies, and the presence of dust
\citep[see the detailed discussion in][]{2008MNRAS.391.1137L},
the presented simplified model qualitatively explains the obtained results of
the relative frequencies of SNe types.

\subsection{Dependence of relative frequencies of SNe types on host activity class}
\label{SNeinactivehosts}

\begin{figure}
  \begin{center}
  \includegraphics[width=0.95\hsize]{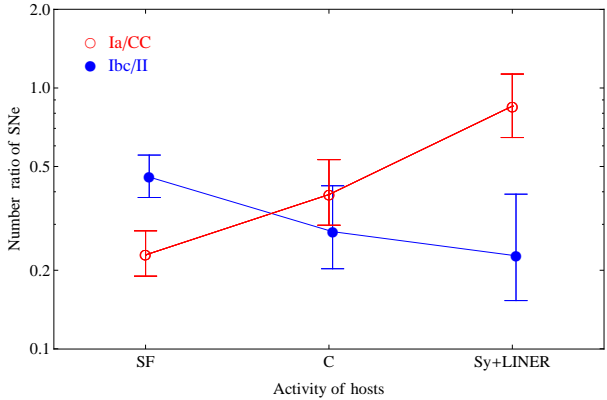}
  \end{center}
  \caption{Relative frequency of SNe types as a function of
           activity classes of their host galaxies.
           The BPT activity classes are available only for 232 hosts with 268 SNe.}
  \label{ratioactiv}
\end{figure}
\begin{table}
  \centering
  \begin{minipage}{84mm}
  \caption{The number ratios of SN types
           for host galaxies with different activity classes.
           The BPT activity classes are available only for 232 hosts
           with 268 SNe.}
  \label{SNratios-activ-table}
    \begin{tabular}{lcccc}
    \hline
  & \multicolumn{1}{c}{SF} & \multicolumn{1}{c}{C} &\multicolumn{1}{c}{Sy$+$LINER} &\multicolumn{1}{c}{All} \\
  \hline
    $N_{\rm Ia}/N_{\rm CC}$ & $0.23_{-0.04}^{+0.05}$ & $0.39_{-0.09}^{+0.14}$ & $0.85_{-0.21}^{+0.28}$ & $0.35_{-0.04}^{+0.05}$ \\
    $N_{\rm Ibc}/N_{\rm II}$ & $0.46_{-0.08}^{+0.10}$ & $0.28_{-0.08}^{+0.14}$ & $0.23_{-0.07}^{+0.16}$ & $0.38_{-0.05}^{+0.07}$ \\
  \hline \\
  \end{tabular}
  \end{minipage}
\end{table}

Given that nuclear activity of galaxies is often correlated
with the star formation in discs of spiral galaxies
\citep[e.g.][]{2003MNRAS.346.1055K,2007ApJS..173..357K,
2005A&A...429..141K,2009AJ....137.3548P},
we examine in this subsection the influence of the nuclear activity
of the hosts on the number ratios of the different SN types.

In Table~\ref{SNratios-activ-table},
for 268 SNe in 232 host galaxies with available BPT activity data,
we calculate the number ratios
by binning the SNe hosts according to their activity class.
The Sy and LINER bins are merged due to an insufficient number of
Type Ibc SNe in the Sy bin.
As in the previous subsections, the number of events is insufficient to evaluate
the $N_{\rm Ic}/N_{\rm Ib}$ ratio in the subsamples with available activity classes.
Fig.~\ref{ratioactiv} presents the relative frequency of
SNe types as a function of activity classes of their host galaxies.

Fig.~\ref{ratioactiv} shows that the $N_{\rm Ia}/N_{\rm CC}$ ratio
increases when moving from SF, C, to Sy$+$LINER classes.
The differences between the $N_{\rm Ia}/N_{\rm CC}$ ratios in
SF versus C ($P_{\rm B} = 0.04$) and C versus Sy$+$LINER ($P_{\rm B} = 0.03$)
subsamples are statistically significant.
In addition, the mean of 1/sSFR for the SNe hosts obtained from the SDSS is
significantly higher for Sy$+$LINER galaxies and lower for SF galaxies.

The $N_{\rm Ibc}/N_{\rm II}$ ratio decreases not significantly when moving from
SF, C, to Sy$+$LINER classes.
The host galaxies with Sy$+$LINER activity classes are
more luminous than the hosts in C bin, which are in turn more luminous
than the hosts in SF subsample:
the difference in the mean of $M_{\rm g}$ between Sy$+$LINER and C subsamples
is about 0.2~mag, while it is about 0.5~mag between the C and SF subsamples.
A KS test yields probabilities of 0.05 and $4 \times 10^{-3}$, respectively,
that the luminosities are drawn from the same distribution.
The trend for luminosity of the hosts can be true also for their metallicities,
however the metallicities extracted from the SDSS
do not allow us to check the mentioned behaviour directly.
Again, we emphasize that any metallicity
constraint from our analyses is unconvincing.

\begin{figure}
  \begin{center}
  \includegraphics[width=0.95\hsize]{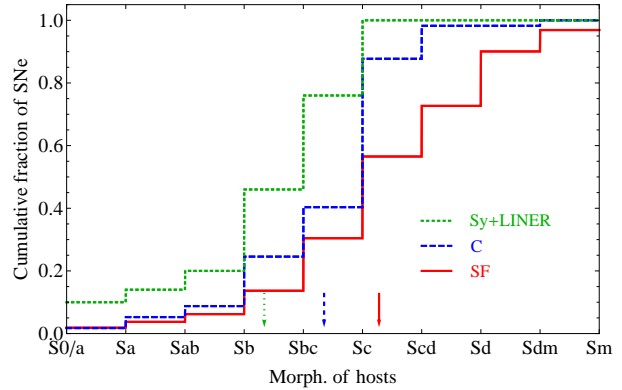}
  \end{center}
  \caption{Cumulative fraction of 268 SNe versus
           host galaxy morphological type. The mean morphologies of
           host galaxies of each activity class of nucleus are shown by arrows.}
  \label{hostactivcum}
\end{figure}
\begin{table}
  \centering
  \begin{minipage}{83mm}
  \caption{KS test probabilities of consistency
           for the distributions of hosts morphologies
           among the different subsamples of activity classes.
           The \emph{t}-type mean values and errors of
           the means are presented in parentheses.}
  \tabcolsep 2.5pt
  \label{hosts_activ_P_table}
    \begin{tabular}{lcllcl}
    \hline
  \multicolumn{2}{c}{Subsample~1} & & \multicolumn{2}{c}{Subsample~2} & \multicolumn{1}{c}{$P_{\rm KS}$} \\
  \hline
    SF & ($5.3\pm0.1$) & versus & C & ($4.4\pm0.2$) & $6 \times 10^{-4}$ \\
    C & ($4.4\pm0.2$) & versus & Sy$+$LINER & ($3.3\pm0.2$) & $2 \times 10^{-3}$ \\
    Sy$+$LINER & ($3.3\pm0.2$) & versus & SF & ($5.3\pm0.1$) & $3 \times 10^{-7}$ \\
  \hline \\
  \end{tabular}
  \end{minipage}
\end{table}

As in previous subsections,
Fig.~\ref{hostactivcum} presents the cumulative distributions of morphologies of
the hosts of 268 SNe according to their activity classes.
The mean morphological type of Sy$+$LINER hosts is earlier by nearly one type than that
of C class of hosts, whose mean morphology is in turn earlier than that of SF
hosts by also nearly one type (see \emph{t}-types in Table~\ref{hosts_activ_P_table}
and arrows in Fig.~\ref{hostactivcum}).
In our sample, only $20\pm10$ per cent of Sy$+$LINER hosts have a level of
disturbance different from normal, whereas about $40\pm8$ per cent of hosts in each C and SF
bins have disturbed morphology.

The behaviour of the number ratios of SNe types
can be explained by the existence of
a time delay between the interaction and the setting of the different classes
of activity: an interaction induces first the SF, then C, and
finally the AGN (Sy$+$LINER) stages of galaxies
\citep[e.g.][]{2013MNRAS.430..638S}.
In this scenario, the interaction is responsible for the morphological
disturbance and inflow of gas towards the centre, which first triggers star formation
\citep[e.g.][]{2001ApJ...559..147S,2013MNRAS.430..638S}
and increases the sSFR \citep[e.g.][]{2008MNRAS.391.1137L}.
Therefore, in the SF stage, we observe a lower value of the $N_{\rm Ia}/N_{\rm CC}$ ratio
and at the same time a somewhat higher value of the $N_{\rm Ibc}/N_{\rm II}$ ratio
as in the morphologically disturbed (interacting/merging) subsample of late-type galaxies
(see Tables~\ref{SNratios-morph-table_s}, \ref{SNratios-morph-table}, and
\ref{SNratios-disturbance-table}).
The starburst then fades with time and the C (composite of SF and AGN) class
evolves to the AGN (Sy$+$LINER) class
\citep[time-scale $\sim0.3-0.5$~Gyr;][]{2010MNRAS.405..933W}
with a comparatively relaxed disturbance,
early-type morphology, poor gas fraction, and old stellar population.
Therefore, in the AGN stage, we observe inverse
values of the ratios as in morphologically
less disturbed (relaxed) early-type galaxies
(see Tables~\ref{SNratios-morph-table_s}, \ref{SNratios-morph-table}, and
\ref{SNratios-disturbance-table}).

Again, we suggest, that even if not considering the above-mentioned scenario,
the morphologies, in combination with the disturbance levels,
can be considered as the most important parameter shaping the number ratios of
the different types of SNe in host galaxies with various classes of activity.

\subsection{Selection effects and possible sample biases}

There are various selection effects and observational biases
that might potentially affect our study:
SN spectroscopic type, host-galaxy morphology and
magnitude biases depending on distance, inclination effects of
the host galaxy disc, Shaw effect depending on the radial distance of
SNe in galaxies, inclusion of uncertain, peculiar or transient types of SNe into the study,
methods of SNe discovery (photographic or CCD imaging),
the SDSS spectroscopic fibre bias, etc.
For more details, the reader is referred to \citetalias{2012A&A...544A..81H},
where many of these effects are discussed.

To check the impact of the mentioned effects on the results of our study,
we repeated the analyses presented in previous subsections using different subsamples of
SNe and their host galaxies, by dividing them according to the galaxy distance,
or selecting only SNe discovered since the use of CCDs (after 2000),
or analysing only SNe which exploded far from the circumnuclear
regions of host galaxies, or considering only
SNe without uncertain, peculiar or transient types.
There are only one
superluminous SN \citep[from the sample of][]{2013MNRAS.431..912Q}, four peculiar
calcium-rich transients \citep[from the sample of][]{2013MNRAS.432.1680Y}, and six peculiar
Type Iax SNe \citep[from the sample of][]{2013ApJ...767...57F}.
We note that our subsample of Type II SNe includes 33 Type IIb and 43 Type IIn SNe.
They have the same distribution according to the host morphology as Type II SNe
(see also table~5 of \citetalias{2012A&A...544A..81H}).
Despite the smaller sizes and larger error bars of the number ratios of SNe types
in the constructed subsamples, the various trends of the number ratios
remain the same qualitatively.

Note that we may have missed weak bars because of inclination
effects, or that in some cases the SDSS images of the hosts may be
too shallow and insufficiently resolved
to detect bars (see also \citetalias{2012A&A...544A..81H}).
For instance, among the spiral host galaxies with inclinations $i \leqslant 50^{\circ}$,
the average bar fraction is $40 \pm 5$ per cent, whereas for hosts without
inclination limit, the average bar fraction is $30 \pm 3$ per cent.
By selecting only galaxies with inclinations $i \leqslant 50^{\circ}$,
we verify that this effect does not bias the trends of the number ratios
presented in Section~\ref{SNeindarredhosts}.

Similarly, the restriction of the SNe hosts with
activity classes to inclinations $i \leqslant 50^{\circ}$ gives us the possibility to
avoid contamination in determination of the activity classes of nuclei, namely
when the circumnuclear regions of galaxies are veiled by
part of a dense and opaque disc.
This effect is weak in our sample and
does not qualitatively affect the trends presented
in Section~\ref{SNeinactivehosts}.

It is also important to note that the observed numbers and
volumetric rates of SNe may be somewhat different from those
predicted from the cosmic SFR (see discussions in
\citealt{2010ApJ...723..329H} and \citealt{2011ApJ...738..154H}).
However, the mentioned difference may affect just the overall scales of the number ratios,
but not affect any comparisons between the number ratios in different galaxy subsamples.

Summarizing, we conclude that there are no strong selection effects
and biases within our SNe and host galaxies samples,
which could drive the observed behaviours of the relative frequencies
of SNe types in the spiral host galaxies presented in this study.

\section{Conclusions}
\label{concl}

In this second paper of a series,
using a well-defined and homogeneous sample
of host galaxies from the SDSS, presented in the first paper
\citep[][]{2012A&A...544A..81H},
we analyse the number ratios of different SN types
($N_{\rm Ia}/N_{\rm CC}$, $N_{\rm Ibc}/N_{\rm II}$, and
$N_{\rm Ic}/N_{\rm Ib}$) in spirals with various morphologies
and in barred or unbarred galaxies.
We also explore the variations in the number ratios
with different levels of morphological disturbance of the hosts.
Our sample consists of 608 spiral galaxies, which host 692 SNe in total.
In addition, we perform a statistical study of 268 SNe discovered in 232 galaxies
with available activity classes of nucleus (SF, C, and Sy$+$LINER).
We propose that the underlying mechanisms
shaping the number ratios of SNe types can be interpreted
within the framework of interaction-induced star formation,
in addition to the known relations between morphologies and stellar populations.

The results obtained in this article are summarized below,
along with their interpretations.

\begin{enumerate}
\item We find a strong trend in the behaviour of $N_{\rm Ia}/N_{\rm CC}$ depending on
      host-galaxy morphological type, such that early-type (high-mass or high-luminosity)
      spirals include proportionally more Type Ia SNe.
      In addition, there is a strong trend in
      the distribution of 1/sSFR, such that sSFR of host galaxies systematically increases
      from early- to late-type spirals.
      The behaviour of $N_{\rm Ia}/N_{\rm CC}$ versus morphology is a simple
      reflection of the behaviour of 1/sSFR versus morphological types of galaxies.
\item The $N_{\rm Ibc}/N_{\rm II}$ ratio is higher in a broad bin of
      early-type hosts.
      The $N_{\rm Ibc}/N_{\rm II}$ distribution is consistent within errors with
      the metallicity distribution of the host galaxies,
      and shows a mild variation with morphology
      of the hosts. In addition, the $N_{\rm Ibc}/N_{\rm II}$ and $N_{\rm Ic}/N_{\rm Ib}$ ratios are
      higher, not significantly, in more luminous (metal-rich) host galaxies.
      However, any metallicity constraint
      from our analyses is very weak.
\item The mean morphological type of spiral galaxies hosting Type Ia SNe is significantly earlier
      than the mean host type for all other types of CC SNe,
      which are, in contrast, consistent with one another.
\item There is no difference between the $N_{\rm Ia}/N_{\rm CC}$ ratios in the subsamples of unbarred
      and barred spirals. However, the $N_{\rm Ibc}/N_{\rm II}$ ratio is higher, with barely
      significance, in unbarred hosts in comparison with the same ratio in barred hosts.
      The number ratios of SNe in barred galaxies are caused by the
      bimodal distribution (two distinct types of bars) of these galaxies with morphology.
      We find that in an individual morphological bin, there is no any significant difference in
      the various number ratios between the unbarred and barred hosts.
\item The $N_{\rm Ia}/N_{\rm CC}$ ratio is nearly constant when changing
      from normal, perturbed to interacting galaxies, then declines in merging galaxies, whereas it
      jumps to the highest value in post-merging/remnant host galaxies.
      During the relatively short time-scale of the merging stage,
      the spiral, gas-rich galaxies do not have enough time to produce many Type Ia SNe,
      but can intensively produce CC SNe, assuming short lifetimes for the CC SNe progenitors.
      In the post-merging/remnant galaxies with longer time-scale,
      the SFRs and morphologies of host galaxies are strongly affected,
      significantly increasing the $N_{\rm Ia}/N_{\rm CC}$ ratio.
\item The $N_{\rm Ibc}/N_{\rm II}$ ratio is nearly constant when changing
      from normal, perturbed to interacting galaxies,
      then jumps to the highest value in merging galaxies
      and slightly declines in post-merging/remnant subsample.
      In our merging hosts, the positions of CC SNe, particularly SNe of Ibc type, mostly coincide
      with the circumnuclear regions and only in few cases with bright H~{\footnotesize II} regions, which is in agreement
      with the previously found central excess of CC SNe in extremely disturbed or merging galaxies.
\item The $N_{\rm Ia}/N_{\rm CC}$ ($N_{\rm Ibc}/N_{\rm II}$) ratio increases (decreases)
      when moving from SF, C, to Sy$+$LINER activity classes (BPT) for the host galaxies. In the invoked scenario,
      the interaction is responsible for morphological disturbances and for partially sending gas inward,
      which first triggers star formation and increases sSFR. Therefore, in the SF stage, we observe a lower value of
      the $N_{\rm Ia}/N_{\rm CC}$ ratio and a somewhat higher value of
      the $N_{\rm Ibc}/N_{\rm II}$ ratio as in morphologically disturbed (interacting or merging)
      late-type galaxies. The starburst then fades with time and the C (composite of SF and AGN) class evolves to
      the AGN (Sy$+$LINER) with a comparatively relaxed disturbance, early-type morphology, poor gas fraction,
      and old stellar population. Therefore,
      in the AGN stage, we observe inverse values of the ratios as in morphologically
      less disturbed (relaxed) early-type galaxies.
\end{enumerate}

We are not able to discriminate between the natures of
different CC SNe progenitors because the time-scales of both the interaction stages and
activity classes of nucleus are much longer than the upper limits of progenitor
lifetimes of Types Ibc and II SNe discussed in the Introduction.
In the forthcoming third paper of this series (Aramyan et al. in preparation),
we study the distributions of the different types of SNe
relative to the spiral arms of the host galaxies with various morphologies,
in order to understand the relationship between the nature of SNe progenitors
and arm-induced star formation.

\section*{Acknowledgements}

AAH, ARP, and LSA acknowledge the hospitality of the
Institut d'Astrophysique de Paris (France) during their
stay as visiting scientists supported by the Collaborative
Bilateral Research Project of the State Committee of Science (SCS)
of the Republic of Armenia and the French
Centre National de la Recherch\'{e} Scientifique (CNRS).
This work was supported by State Committee Science MES RA,
in frame of the research project number SCS~13--1C013.
AAH is also partially supported by the ICTP.
VZA and JMG are supported by grants SFRH/BPD/70574/2010 and
SFRH/BPD/66958/2009 from FCT (Portugal), respectively.
VZA would further like to thank for the support by the ERC under
the FP7/EC through a Starting Grant agreement number 239953.
DK acknowledges financial support from the Centre
National d'\'{E}tudes Spatiales (CNES).
MT is partially supported by the PRIN-INAF 2011 with the project
Transient Universe: from ESO Large to PESSTO.
This work was made possible in part by a research grant from the
Armenian National Science and Education Fund (ANSEF)
based in New York, USA.
Finally, we are especially grateful
to our referee for his/her constructive comments.
Funding for SDSS-III has been provided by the Alfred P.~Sloan Foundation,
the Participating Institutions, the National Science Foundation,
and the US Department of Energy Office of Science.
The SDSS--III web site is \href{http://www.sdss3.org/}{http://www.sdss3.org/}.
SDSS--III is managed by the Astrophysical Research Consortium for the
Participating Institutions of the SDSS--III Collaboration including the
University of Arizona, the Brazilian Participation Group,
Brookhaven National Laboratory, University of Cambridge,
University of Florida, the French Participation Group,
the German Participation Group, the Instituto de Astrofisica de Canarias,
the Michigan State/Notre Dame/JINA Participation Group,
Johns Hopkins University, Lawrence Berkeley National Laboratory,
Max Planck Institute for Astrophysics, New Mexico State University,
New York University, Ohio State University, Pennsylvania State University,
University of Portsmouth, Princeton University, the Spanish Participation Group,
University of Tokyo, University of Utah, Vanderbilt University,
University of Virginia, University of Washington, and Yale University.

\bibliography{snbibII}

\section*{Supporting information}

Additional Supporting Information may be found in
the online version of this article:\\
\\
(\href{http://mnras.oxfordjournals.org/lookup/suppl/doi:10.1093/mnras/stu1598/-/DC1}
{http://mnras.oxfordjournals.org/lookup/suppl/doi:10.1093/mnras/}
\href{http://mnras.oxfordjournals.org/lookup/suppl/doi:10.1093/mnras/stu1598/-/DC1}
{stu1598/-/DC1}).

\label{lastpage}

\end{document}